\documentclass[12pt,preprint]{aastex}

\begin{document}
\shorttitle{Spectral Evolution in the Steep Decay Phase}
\shortauthors{Lin et al.}
\title{Steep Decay Phase Shaped by the Curvature Effect. II. Spectral Evolution}
\author{Da-Bin Lin\altaffilmark{1,2}, Hui-Jun Mu\altaffilmark{1,2,3}, Yun-Feng Liang\altaffilmark{4}, Tong Liu\altaffilmark{3}, Wei-Min Gu\altaffilmark{3}, Rui-Jing Lu\altaffilmark{1,2}, Xiang-Gao, Wang\altaffilmark{1,2}, and En-Wei Liang\altaffilmark{1,2}
}
\altaffiltext{1}{GXU-NAOC Center for Astrophysics and Space Sciences, Department of Physics, Guangxi University, Nanning 530004, China; lindabin@gxu.edu.cn, muhuijun@stu.xmu.edu.cn}
\altaffiltext{2}{Guangxi Key Laboratory for Relativistic Astrophysics, the Department of Physics, Guangxi University, Nanning 530004, China}
\altaffiltext{3}{Department of Astronomy, Xiamen University, Xiamen, Fujian 361005, China}
\altaffiltext{4}{Purple Mountain Observatory, Chinese Academy of Sciences, Nanjing 210008, China}
\begin{abstract}
We derive a simple analytical formula to describe the evolution of spectral index
$\beta$ in the steep decay phase shaped by the curvature effect
with assumption that the spectral parameters and Lorentz factor of jet shell is the same for different latitude.
Here, the value of $\beta$ is estimated in 0.3$-$10keV energy band.
For a spherical thin shell with a cutoff power law (CPL) intrinsic radiation spectrum,
the spectral evolution can be read as a linear function of observer time.
For the situation with Band function intrinsic radiation spectrum,
the spectral evolution may be complex.
If the observed break energy of radiation spectrum is larger than 10keV,
the spectral evolution is the same as that shaped by jet shells with a CPL spectrum.
If the observed break energy is less than 0.3keV, the value of $\beta$ would be a constant.
Others, the spectral evolution can be approximated as a logarithmal function of the observer time in generally.
\end{abstract}
\section{Introduction}\label{Sec:Introduction}
Gamma-ray bursts (GRBs) are the most powerful electromagnetic explosions in the Universe.
The so-called $\gamma$-ray prompt emission phase,
which always triggers the observation of Burst Alert Telescope (BAT, \citealp{Barthelmy2005SSRv}),
exhibits highly variable and diverse morphologies.
The highly variabilities in this phase may originate from the central engine activities
(e.g., \citealp{Ouyed2003,Proga2003,Lei2007,Liu2010,Lin2016MN,Zhang2016}),
the processes during jet propagation (\citealp{Aloy2002}; \citealp{Morsony2007}; \citealp{Morsony2010}),
and the relativistic motion (e.g., mini-jets/turbulence) in the emission region
(\citealp{Lyutikov2003}; \citealp{Yamazaki2004}; \citealp{Kumar2009}; \citealp{Lazar2009};
\citealp{Narayan2009}; \citealp{Lin2013}; \citealp{ZhangBo2014}).
Following the prompt emission is a smooth steep decay phase,
which is always observed at $\sim 10^2-10^3$ seconds after the burst trigger and in the X-ray band
(\citealp{Vaughan2006,Cusumano2006, OBrien2006}).
By extrapolating the prompt $\gamma$-ray light curve to X-ray band,
the smooth steep decay phase can connect to this extrapolated X-ray light curve smoothly.
Thus, it is believed that the steep decay phase may be the ``tail'' of the prompt emission
(\citealp{Barthelmy2005, OBrien2006, Liang2006}).
The steep decay phase is also observed in the decay phase of flares (e.g., \citealp{Jia2015}; \citealp{Mu2016}; \citealp{Uhm_ZL-2016-Zhang_B}).

For the steep decay phase, the temporal decay index $\alpha$ of the observed flux
($F_E\propto (t_{\rm obs}^{\rm tg})^{-\alpha}$) is found to be correlated with the spectral index $\beta$,
where $t_{\rm obs}^{\rm tg}$ is the observer time after the trigger.
This led to the development of the ``curvature effect'' model
(\citealp{Zhang2006,Liang2006,Wu2006,Yamazaki2006}).
When emission in a spherical relativistic jet shell ceases abruptly,
the observed flux is controlled by high latitude's emission of the jet shell.
In this situation, the photons from higher latitude would be observed later and have a lower Doppler factor.
Then, the observed flux would progressively decrease.
For a power-law radiation spectrum ($F'\propto E'^{-\beta}$ with $\beta=\rm constant$) in the jet shell comoving frame,
the relation between $\alpha$ and $\beta$ can be found, i.e., $\alpha=2+\beta$
(see \citealp{Uhm2015} for details; \citealp{Kumar2000,Dermer2004,Dyks2005}).
As shown in \cite{Nousek2006}, above relation is in rough agreement
with the data on the steep decay phase of some \emph{Swift} bursts.
Adopting a time-averaged $\beta$ in the steep decay phases,
\cite{Liang2006} finds that $\alpha=2+\beta$ is generally valid.
\cite{Lin_DB-2017-Mu_HJ} points out that $\alpha=2+\beta$
is only valid for a power-law radiation spectrum.

The strong evolution of spectral index $\beta$ is always found in the steep decay phase
(\citealp{Zhang2007,Butler2007,Starling2008,Zhang2009,Mu2016}).
Several effects have been done to explain the strong $\beta$ evolution in the scenario of curvature effect.
Since photons observed later may be from the higher latitude angle ($\theta$) of jet shell,
the softening spectrum may be due to a $\theta$-dependent spectral shape in the comoving frame of the jet shell (\citealp{Zhang2007}).
In this scenario, the emitting region with the hardest spectrum
in the jet shell should be always in the direction of $\theta = 0$
for most of steep decay phases with strong spectral evolution (e.g., \citealp{Mu2016}).
It seems to be contrived.
Besides a $\theta$-dependent spectral shape,
the softening spectrum may be due to a non-power-law intrinsic radiation spectrum
with $\theta$-independent spectral parameters (\citealp{Zhang2009}).
Since the radiation from different $\theta$ is observed at different time and with different Doppler factor,
the observed X-ray emission would be from different segments of the non-power-law intrinsic spectrum.
Then, one would find a strong spectral evolution in the steep decay phases.
This scenario has been studied in \cite{Zhang2009} for the steep decay phase of the prompt emission in GRB~050814.
However, what is the evolution pattern of the spectral index
in the steep decay phase does not discussed in details.
Then, we try to derive an analytical formula to describe the $\beta$ evolution in the steep decay phase
with a non-power-law intrinsic radiation spectrum and $\theta$-independent spectral parameters.

The paper is organized as follows.
Since we try to test our analytical formula of $\beta$ evolution based on the numerical simulations,
the procedure of our numerical simulations is presented in Section~\ref{Sec:Numerical calculation Procedure}.
The functional form of spectral evolution in the steep decay phase is derived
and tested in Sections~\ref{Sec:Evolution of Spectra} and \ref{Sec:Testing}, respectively.
Conclusions and discussion are made in Section~\ref{Sec:conclusion}.

\section{Procedure for Simulating Jet Emission}\label{Sec:Numerical calculation Procedure}
The emission of a spherical thin jet shell with jet opening angle $\theta_{\rm jet}$ radiating from $r_0$ to $r_e$ is our focus,
where $r_0$ and $r_e$ are the location of jet shell estimated with respect to the jet base.
We assume that (1) the central axis of jet shell coincides with the observer's line of sight;
(2) the jet shell has no $\theta$-dependent spectral parameters and Lorentz factor.
\footnote{
If assumption (2) does not hold, the $\beta$ evolution would depend on the viewing angle.
There are several previous works taking into account the viewing angle effect on the observed flux
(e.g., \citealp{Yamazaki_R-2003-Ioka_K}).
For most of steep decay phases with strong spectral evolution,
the value of $\beta$ is the lowest at the beginning of the steep decay phase (e.g., \citealp{Mu2016}; \citealp{Uhm_ZL-2016-Zhang_B}).
If assumption (2) does not hold,
the emitting region with the hardest intrinsic spectrum in the jet shell may be always in the direction of our sight.
It seems to be contrived.
Then, the $\theta$-dependence of the spectral parameters and Lorentz factor may be weak.}
The procedure for simulating jet emission is detailed in \cite{Lin_DB-2017-Mu_HJ}.
In this section, we present a brief description about this model.

We assume the jet shell locating at radius $r$ for time $t$,
where $r$ is measured with respect to the jet base.
For performing the simulations about the jet radiation,
the jet shell is modelled with a number of emitters randomly distributed among the jet shell.
The observed time of photons from an emitter located at ($r, \theta$) is
\begin{equation}\label{Eq:t_obs}
{t_{{\rm{obs}}}} =\left\{ \int_{r_0}^{{r}} {[1 - \beta_{\rm jet}(l)]} \frac{{dl}}{c\beta_{\rm jet}(l) }+ \frac{r(1- \cos \theta)}{c}\right\}(1+z),
\end{equation}
where $c\beta_{\rm jet}(l)=cdr/dt$ is the velocity of jet shell at radius $r=l$, $c$ is the light velocity,
$\theta$ is the polar angle of the emitter with respect to the line of sight in spherical coordinates
(the origin of coordinate is at the jet base),
and $z$ is the redshift of the explosion producing the jet shell.

The radiation mechanism of an emitter is always discussed in the synchrotron process or the inverse Compton process.
In our work, the shape of radiation spectrum is important rather than the detailed radiation processes.
Following the work of \cite{Uhm2015}, the radiation spectrum of an electron with $\gamma'_e\; (>>1)$ is assumed as
\begin{equation}\label{eq:spec_power_single_ensemble}
P'(E') =P'_0\, H'(E'/\hat{E}'_0),
\end{equation}
where $P_0^{\prime}$ describes the spectral power observed in the jet shell comoving frame
and $\hat{E}'_0$ is the characteristic photon energy of electron emission.
Thus, the total radiation power of an emitter in the comoving frame is $n'_eP'_0\, H'(E'/\hat{E}'_0)$,
where $n'_e$ is the total number of electrons with $\gamma'_e$ in an emitter.
For the functional form of $H'(x)$, we study following two cases:
\begin{equation}\label{Eq:H function}
\begin{array}{*{20}{c}}
{{\rm{Case\;(I)}}:}&{H'(x) = \left\{ {\begin{array}{*{20}{c}}
{{x^{\hat \alpha + 1}}\exp ( { - x} ),}&{x \leqslant( {\hat \alpha- \hat \beta} ),}\\
{{{( {\hat \alpha - \hat \beta})}^{\hat \alpha - \hat \beta}}\exp ( {\hat \beta - \hat \alpha} ){x^{\hat \beta + 1}},}&{x \geqslant ( {\hat \alpha - \hat \beta}),}
\end{array}} \right.}\\
{{\rm{Case\;(II)}}:}&{H'(x) = x^{\hat \alpha+1}\exp (-x),}
\end{array}
\end{equation}
where $\hat \alpha$ and $\hat \beta$ are constants.
The spectral shape in Case (I) is the so-called ``Band-function'' spectrum (\citealp{Band1993}).
In some bursts, the observed prompt (or X-ray flare) emission can be fitted with a cutoff power-law (CPL) spectrum, i.e., Case (II).
Then, the spectral evolution for a jet shell with Case (II) is also studied.
A photon in the comoving frame with energy $E'$ is boosted to $E=DE'/(1+z)$ in the observer's frame,
where $D$ is the Doppler factor described as
\begin{eqnarray}\label{}
D ={\left[ {{\Gamma }(1 - {\beta _{{\rm{jet}}}}\cos \theta )} \right]^{ - 1}},
\end{eqnarray}
and $\Gamma$ is the Lorentz factor of the jet shell.
During the shell's expansion for $\delta t\; (\sim 0)$,
the observed spectral energy $\delta U$ from an emitter
into a solid angle $\delta \Omega$ in the direction of the observer is given as (\citealp{Uhm2015})
\begin{equation}\label{Eq:Numerical calculation-Flux}
\delta {U_E} (t_{\rm obs}) = \left({D^2}\delta \Omega\right)\left( \frac{{\delta t}}{\Gamma }\right)\frac{1}{{4\pi }}{{n'}_e}{{P'}_0}H'\left( {\frac{{E(1 + z)}}{{D{{\hat E'}_0}}}} \right),
\end{equation}
where the emission of electrons is assumed isotropically in the jet shell comoving frame (c.f. \citealp{Geng_JJ-2017-Huang_YF}).

The procedures for obtaining the observed flux is shown as follows.
Firstly, an expanding jet is modelled with a series of jet shells
at radius $r_0,\;r_1=r_0+\beta_{\rm jet}(r)c{\delta t},\;r_2=r_1+\beta_{\rm jet}(r_1)c{\delta t},\;\cdot\cdot\cdot,r_n=r_{n-1}+\beta_{\rm jet}(r_{n-1})c{\delta t},\;\cdot\cdot\cdot$
appearing at the time $t=0{\rm s},\;{\delta t},\;{2\delta t},\;\cdot\cdot\cdot, n{\delta t},\;\cdot\cdot\cdot $
with velocity $c\beta_{\rm jet}(r),\;c\beta_{\rm jet}(r_1),\;c\beta_{\rm jet}(r_2),\;\cdot\cdot\cdot,c\beta_{\rm jet}(r_n),\;\cdot\cdot\cdot$, respectively.
During the shell's expansion for $\delta t$,
the shell move from $r_{n-1}$ to $r_n$ with the same radiation behavior for emitters.
Secondly, we produce $N$ emitters centred at ($r_n$, $\theta$, $\varphi$) in spherical coordinates,
where the value of $\cos\theta$ and $\varphi$ are randomly picked up from linear space of $[\cos\theta_{\rm jet},1]$ and $[0,2\pi]$, respectively.
The observed spectral energy from an emitter
during the shell's expansion from $r_{n-1}$ to $r_n$ is calculated with Equation~(\ref{Eq:Numerical calculation-Flux}).
By discretizing the observer time $t_{\rm obs}$ into a series of time intervals,
i.e., $[0, {\delta t_{\rm obs}}],\,[{\delta t_{\rm obs}},2{\delta t_{\rm obs}}]\, \cdot\cdot\cdot\,,[(k-1){\delta t_{\rm obs}}, k{\delta t_{\rm obs}}],\cdot\cdot\cdot$,
we can find the total observed spectral energy
\begin{equation}
\left. U_E \right|_{\left[(k - 1)\delta {t_{{\rm{obs}}}},k\delta {t_{{\rm{obs}}}} \right)}=
\sum\limits_{(k-1)\delta t_{\rm obs}\leqslant t_{\rm obs}<k\delta t_{\rm obs}}{\delta {U_E} (t_{\rm obs})}
\end{equation}
in the time interval $[(k-1){\delta t_{\rm obs}}, k{\delta t_{\rm obs}}]$ based on Equations~(\ref{Eq:t_obs}) and (\ref{Eq:Numerical calculation-Flux}).
Then, the observed flux at the time $(k/2-1){\delta t_{\rm obs}}$ is
\begin{equation}
F_E=\frac{{{{\left. U_E \right|}_{\left[(k - 1)\delta {t_{{\rm{obs}}}},k\delta {t_{{\rm{obs}}}} \right)}}}}{{D_{\rm{L}}^2\delta {t_{{\rm{obs}}}}\delta \Omega}},
\end{equation}
where $D_{\rm L}$ is the luminosity distance of the jet shell with respect to the observer.

In our numerical simulations,
the jet shell is assumed to begin radiation at radius $r=10^{14}\rm cm$
with a Lorentz factor $\Gamma(r)=\Gamma_0=300$ and $\hat{E}'_0(r)=\hat{E}'_{0,r}$.
The evolution of Lorentz factor $\Gamma$ and $\hat{E}'_0$ are assumed as
\begin{equation}
\Gamma(r)=\Gamma_0\left(\frac{r}{r}\right)^s
,\;\;
\hat{E}'_0(r)=\hat{E}'_{0,r}\left(\frac{r}{r}\right)^{w-s}.
\end{equation}
The value of $n'_eP'_0$ is assumed to increase with time $t'$ in the jet comoving frame,
i.e., $n'_eP'_0=n'_{e,0}P'_{0,0}t'$, and $t'=0$ is set at the radius $r$,
where $n'_{e,0}$ and $P'_{0,0}$ are constants.
The value of $N>>1$, $\delta t << t_{c,r}$, $\theta_{\rm jet}>>1/\Gamma_0$, and $\delta t_{\rm obs}=0.005t_{c,r}$
are adopted and remained as constants in a numerical simulation, where $t_{c,r}=r(1+z)/\Gamma^2c$.
By changing the observed photon energy $E$ and running above numerical simulation again,
we can find the observed flux $F_E$ at different $E$.
Since the observational energy band of Swift X-Ray Telescope (XRT) used to estimate the spectral index is
$[0.3{\rm keV}, 10{\rm keV}]$, we obtain the observed flux $F_E$
at photon energy $E=0.3{\rm keV}, 0.3\times 1.12 {\rm keV}, 0.3\times 1.12^2 {\rm keV}, \cdots, 10{\rm keV}$.
With $F_E$ observed at different $E$,
we fit the spectrum with a power-law function $(E/1{\rm keV})^{-\beta}$ to find the value of $\beta$.
The total duration of our light curves are set as $50t_{c,r}$.
Then, the obtained data would be significantly large.
To reduce the file size of our figures, we only plot the data in the time interval with $k$
satisfying $(k-1){\delta t_{\rm obs}}<1.1^m\times 0.01t_{c,r}+t_0<k{\delta t_{\rm obs}}$,
where $m\;(\geqslant 0)$ is an any integer and $t_0$ is the observer time set for $\tilde{t}_{\rm obs}=0$ (see Section~\ref{Sec:Evolution of Spectra}).
The spectral evolution pattern in these figures are the same as those plotted based on all of data from our numerical simulations.

\section{Analytical Formula of Spectral Evolution}\label{Sec:Evolution of Spectra}
We first analyze the spectral evolution for radiation from an extremely fast cooling thin shell (EFCS).
For this situation, we assume the radiation behavior of jet shell unchanged during the shell's expansion time $\delta t$ ($\sim 0$).
Then, we have $r=r_0+\beta_{\rm jet}c\delta t \sim r_0$ and Equation~(\ref{Eq:t_obs}) can be reduced to
\begin{equation}\label{Eq:t_obs_thin}
t_{\rm obs}=(r/c)(1-\cos\theta)(1+z),
\end{equation}
which describes the delay time of photons from ($r,\;\theta$) with respect to those from ($r,\;\theta=0$).
It should be noted that the beginning of the phase shaped by the curvature effect
in this situation ($\delta t\sim 0$) is at around $t_{\rm obs}= 0$.
With $\Gamma\gg 1$, $D$ can be reduced to
\begin{equation}
D \approx {\left\{ {\Gamma  - \Gamma \left( {1 - \frac{1}{2\Gamma ^2}} \right)\left [1 - (1 - \cos \theta )\right ]} \right\}^{ - 1}} \approx {\left[ {\frac{1}{{2\Gamma }} + \Gamma (1 - \cos \theta )} \right]^{ - 1}},
\end{equation}
or
\begin{equation}\label{Eq:D}
D\approx \frac{2\Gamma}{1+t_{\rm obs}/t_{c,r}},
\end{equation}
where $t_{c,r}$ is the characteristic timescale of shell curvature effect at radius $r$,
\begin{equation}\label{Eq:Gamma}
t_{c,r}=\frac{r(1 + z)}{2\Gamma^2c}.
\end{equation}
The difference between $D$ and ${2\Gamma}/{(1+t_{\rm obs}/t_{c,r})}$ can be neglected for significantly large value of $\Gamma$. Then, we use $D={2\Gamma}/{(1+t_{\rm obs}/t_{c,r})}$ in our analysis.

For the observer time interval $\delta t_{\rm obs}$,
the observed total number of emitter is $N|\delta (\cos\theta)|/(1-\cos\theta_{\rm jet})$
with $|\delta (\cos\theta)|=c\delta t_{\rm obs}/r(1+z)$ derived based on Equation~(\ref{Eq:t_obs_thin}),
where $\theta_{\rm jet}$ is the jet opening angle.
Then, the observed flux at the time $t_{\rm obs}$ is
\begin{equation}
F_E=\frac{\delta {U_E} (t_{\rm obs})N|\delta (\cos\theta)|/(1-\cos\theta_{\rm jet})}{D_{\rm{L}}^2\delta t_{\rm obs}\delta \Omega},
\end{equation}
or,
\begin{equation}\label{Eq:Spectral_Evolution}
F_E\propto D^2H'(E(1+z)/D\hat E'_0)/\Gamma.
\end{equation}
The method used to derive Equation~(\ref{Eq:Spectral_Evolution}) is from \cite{Uhm2015}.
The reader can read the above paper for the details.

The observed spectral index $\beta$ is always estimated with XRT observations.
The observational energy band of XRT is $[0.3{\rm keV}, 10{\rm keV}]$.
Then, $\beta$ can be approximately described as
\begin{equation}\label{Eq:Beta_0.3-10}
\beta \approx \beta_{\rm es}=-\frac{\log({F_{10\rm keV}}/{F_{0.3\rm keV}})}{\log({10\rm keV}/{0.3\rm keV})}.
\end{equation}
For Case (II), we have
\begin{equation}\label{Eq:Case_II_0}
\beta_{\rm es} = { - \hat \alpha  - 1{\rm{ + }}\frac{{10{\rm{keV}} - {\rm{0}}{\rm{.3keV}}}}{{{E}_{0,r}\left[ {\ln \left( {10{\rm{keV}}} /{{\rm{0}}{\rm{.3keV}}} \right)} \right]}}},
\end{equation}
where $E_{0,r}$ is read as
\begin{equation}\label{Eq:E_0_r}
E_{0,r}(t_{\rm obs})=\frac{\hat{E}'_0(r)D}{1+z}=\frac{2\Gamma\hat{E}'_0(r)}{(1+t_{\rm obs}/t_{c,r})(1+z)}.
\end{equation}
For Case (I), however, the relation of $\beta_{\rm es}$ and $E_{0,r}$ may be different for different value of $E_{0,r}(t_{\rm obs})$.\\
For Case (I) with $10{\rm{keV}} \leqslant( {\hat \alpha  - \hat \beta })E_{0,r}$,
we have
\begin{equation}\label{Eq:Case_II}
\beta_{\rm es} = { - \hat \alpha  - 1{\rm{ + }}\frac{{10{\rm{keV}} - {\rm{0}}{\rm{.3keV}}}}{{{E}_{0,r}\left[ {\ln \left( {10{\rm{keV}}} /{{\rm{0}}{\rm{.3keV}}} \right)} \right]}}},
\end{equation}
which is the same as Equation~(\ref{Eq:Case_II_0}).\\
For Case (I) with $0.3{\rm{keV}} \geqslant ( {\hat \alpha  - \hat \beta } ){E}_{0,r}$, one can find
\begin{equation}\label{Eq:beta_up}
\beta_{\rm es}=- \hat{\beta} - 1.
\end{equation}
For Case (I) with $0.3{\rm{keV}}< (\hat \alpha  - \hat \beta ){E}_{0,r} < 10{\rm{keV}}$, we have
\[\beta_{\rm es} =  - \frac{{\log\left({{F_{10{\rm{keV}}}}/F_{E_{0,r}}}\right) + \log{{(F_{E_{0,r}}}}/{{F_{{\rm{0}}{\rm{.3keV}}}}} )}}{{\log( {10{\rm{keV}}}/ {{\rm{0}}{\rm{.3keV}}})}}\]
\begin{equation}
=- \frac{{\log ( {10{\rm{keV}}} / {E_{0,r}} )}}{{\log ( {10{\rm{keV}}} / {{\rm{0}}{\rm{.3keV}}} )}}\frac{{\log ( {{F_{10{\rm{keV}}}}} / {{F_{E_{0,r}}}} )}}{{\log ( {10{\rm{keV}}} / {E_{0,r}} )}}
- \frac{{\log ( {E_{0,r}} / {{\rm{0}}{\rm{.3keV}}} )}}{{\log ( {10{\rm{keV}}} / {{\rm{0}}{\rm{.3keV}}} )}}\frac{{\log ( {{F_{E_{0,r}}}} / {{F_{{\rm{0}}{\rm{.3keV}}}}} )}}{{\log ( {E_{0,r}} / {{\rm{0}}{\rm{.3keV}}} )}},
\end{equation}
or,
\begin{equation}\label{Eq:beta_m}
\beta_{\rm es} ={ - (\hat \beta  + 1)\frac{{\ln (10{\rm{keV}}/E_{0,r})}}{\ln (10/0.3)} - (\hat \alpha  + 1)\frac{\ln (E_{0,r}/0.3{\rm keV})}{\ln (10/0.3)} + \frac{E_{0,r} - 0.{\rm 3keV}}{E_{0,r}\ln (10/0.3)}}.
\end{equation}
For Case (I), we compare the value of $\beta$ and $\beta_{\rm es}$ in
the upper-left panel of Figure~1 by changing the value of $E_{0,r}$,
where the value of $\hat \alpha =-1$ and $\hat \beta=-2.3$ are adopted.
The value of $\beta$ is obtained by fitting $F_{0.3{\rm keV}}, F_{0.3\times 1.12 {\rm keV}},\cdots, F_{10{\rm keV}}$
with $(E/1{\rm keV})^{-\beta}$.
In this panel, the value of $\beta$ and $\beta_{\rm es}$ are shown with black ``$+$'' and blue dashed line, respectively.
The value of $\beta_{\rm es}$ presents a well estimation about the spectral index $\beta$.
However, the deviation of $\beta_{\rm es}$ with respect to $\beta$ can be easily found for $0.3{\rm keV} \lesssim E_{0,r} \lesssim  20{\rm keV}$.

Then, we would like to present a better estimation ($\beta'_{\rm es}$) about the value of $\beta$.
For Case (II) or Case (I) with $E_{0,r}\geqslant 10{\rm{keV}}/( {\hat \alpha  - \hat \beta } )$,
the deviation of $\beta_{\rm es}$ relative to $\beta$
is owing to that we use a power-law function to fit a cutoff power-law spectrum.
In Figure~1, one can easily find the behavior of $\beta_{\rm es}>\beta$.
Then, we adopt
\begin{equation}\label{Eq:exp_cut_off}
\beta'_{\rm es} = { - \hat \alpha  - 1+ 0.857\frac{{10{\rm{keV}} - {\rm{0}}{\rm{.3keV}}}}{{E_{0,r}\left[ {\ln \left( {10{\rm{keV}}} / {{\rm{0}}{\rm{.3keV}}} \right)} \right]}}}.
\end{equation}
to estimate the spectral index $\beta$.\\
For Case (I) with $E_{0,r} \leqslant 0.3{\rm{keV}} /( {\hat \alpha  - \hat \beta } )$, we adopt
\begin{equation}\label{Eq:beta_up}
\beta'_{\rm es}=- \hat{\beta} - 1.
\end{equation}
For Case (I) with $0.3{\rm{keV}}< (\hat \alpha  - \hat \beta )E_{0,r} < 10{\rm{keV}}$,
we have
\begin{equation}\label{Eq:beta_m}
\beta_{\rm es} \propto
\frac{(\hat \beta  - \hat \alpha )\ln(E_{0,r}/1{\rm keV})}{{\ln (10/0.3)}} - \frac{{{\rm{0}}.{\rm{3keV}}}}{{E_{0,r} \ln ( 10/0.3 )}}.
\end{equation}
Then, the following form
\begin{equation}\label{Eq:beta_m}
\beta'_{\rm es} \propto
x\frac{(\hat \beta  - \hat \alpha )\ln(E_{0,r}/1{\rm keV})}{{\ln (10/0.3)}} - y\frac{{{\rm{0}}.{\rm{3keV}}}}{{E_{0,r} \ln ( 10/0.3 )}}
\end{equation}
is used to estimate the spectral index $\beta$.
Here, the value of $x$ and $y$ are obtained by requiring the continuity of $\beta'_{\rm es}$
at $E_{0,r}=0.3{\rm{keV}}/( {\hat \alpha  - \hat \beta } )$ and $10{\rm{keV}}/(\hat \alpha -\hat \beta )$,
i.e.,
\begin{equation}
0.2371 + x - 0.2766y - 1 = 0.
\end{equation}
Then, we adopt $x=1.22$ and $y=1.65$ in this work, i.e.,
\begin{equation}\label{Eq:beta_m}
\beta '_{\rm es} =  A+
1.22\frac{(\hat \beta  - \hat \alpha )\ln(E_{0,r}/1{\rm keV})}{{\ln (10/0.3)}} - 1.65\frac{{{\rm{0}}.{\rm{3keV}}}}{{E_{0,r} \ln ( 10/0.3 )}}
\end{equation}
being used for the situations with Case (I) and $0.3{\rm{keV}}< (\hat \alpha  - \hat \beta )E_{0,r} < 10{\rm{keV}}$,
where $A=- 1 + 0.0573\hat \alpha  - 1.0573\hat \beta  + 0.3479(\hat \beta  - \hat \alpha )\ln (\hat \alpha  - \hat \beta )$.
The value of $\beta'_{\rm es}$ for situations with Case (I) and different $E_{0,r}$ can be found in Figure~1 with red solid lines.
One can find that the deviation of $\beta'_{\rm es}$ relative to $\beta$ is very small
for an EFCS with Case (I).
Then, we use $\beta'_{\rm es}$ to describe the spectral index for situations with Case (I).
In addition, Equation~(\ref{Eq:exp_cut_off}) is adopted to describe the spectral index for situations with Case (II).
In Figure~1, one can also find that the value of $(\beta+\hat\alpha+1)/(\hat\alpha-\hat\beta)$ with respect to $E_{0,r}$
is almost the same for different $\hat\alpha$ and $\hat\beta$.
This reveals that the evolution pattern of spectral index would be almost the same for different $\hat\alpha$ and $\hat\beta$.
Then, we only discuss Case (I) with $\hat \alpha =-1$ and $\hat \beta=-2.3$.

By substituting Equation~(\ref{Eq:E_0_r}) into $\beta'_{\rm es}$,
the analytical formula of $t_{\rm obs}$-dependent $\beta$ can be obtained, i.e.,
\begin{equation}\label{Eq:Spectral Evolution_ini}
\begin{array}{*{20}{c}}
{{\rm{Case\;(I)}}:}&{\beta(t_{\rm obs}) = \left\{ {\begin{array}{*{20}{c}}
{- \hat \alpha- 1 + 7.9\kappa ( {1 + {t_{\rm obs}}/{t_{c,r}}} ),}&{10{\rm{keV}} \leqslant( {\hat \alpha  - \hat \beta })E_{0,r}(t_{\rm obs}),}\\
{a + b{t_{{\rm{obs}}}}/{t_{c,r}} + c\ln ( {1 + {t_{{\rm{obs}}}}/{t_{c,r}}}  ),}&{0.3{\rm{keV}}< (\hat \alpha  - \hat \beta ){E_{0,r}(t_{\rm obs})} < 10{\rm{keV}},}\\
{-  \hat \beta  - 1},&{0.3{\rm{keV}} \geqslant ( {\hat \alpha  - \hat \beta } ){E_{0,r}(t_{\rm obs})}},
\end{array}} \right.}\\
{{\rm{Case\;(II)}}:}&{\beta(t_{\rm obs}) = - \hat \alpha- 1 + 7.9\kappa ( {1 + {t_{\rm obs}}/{t_{c,r}}} ),}
\end{array}
\end{equation}
where $\kappa$, $a$, $b$, and $c$ are defined as follows:
\begin{equation}
\kappa  = \frac{0.3{\rm keV}}{E_{0,r}(t_{\rm obs}=0)},
\end{equation}
\begin{equation}
a=- 1 + 0.4762\hat \alpha- 1.4762 \hat \beta  + 0.3479(\hat \beta  - \hat \alpha )\ln (\hat \alpha  - \hat \beta ) + 0.3479(\hat \alpha  - \hat \beta )\ln \kappa+b,
\end{equation}
\begin{equation}
b =  - 1.65\frac{0.3{\rm keV}}{E_{0,r}(t_{\rm obs}=0)\ln(10/0.3)} =  - 0.4706\kappa,
\end{equation}
\begin{equation}
c = 1.22\frac{\hat \alpha  - \hat \beta}{\ln ( {10/0.3})} = 0.3479(\hat \alpha  - \hat \beta ).
\end{equation}
For situations with Case (I),
$({\hat \alpha  - \hat \beta } )E_{0,r}(t_{\rm obs}=0)$ is the break energy of Band function observed at $t_{\rm obs}=0$;
for situations with Case (II),
$E_{0,r}(t_{\rm obs}=0)$ is the cutoff energy of CPL spectrum observed at $t_{\rm obs}=0$.
In practice, we may be interested on the steep decay phase with $t_{\rm obs}\geqslant t_0\, (\geqslant 0)$.
By defining $\tilde{t}_{\rm obs}=t_{\rm obs}-t_0$,
Equation~(\ref{Eq:D}) is reduced to
\begin{equation}\label{Eq:D_t0}
D(\tilde{t}_{\rm obs})=\frac{2\Gamma}{1+t_0/t_{c,r}}\frac{1}{1+\tilde{t}_{\rm obs}/(t_{c,r}+t_0)}
=\frac{D_{t_0}}{1+\tilde{t}_{\rm obs}/\tilde{t}_{c,r}},
\end{equation}
where $D_{t_0}={2\Gamma}/({1+t_0/t_{c,r}})$ is the Doppler factor of emitter observed at $\tilde{t}_{\rm obs}=0$ (or $t_{\rm obs}=t_0$)
and $\tilde{t}_{c,r}=t_{c,r}+t_0$ is adopted.
With Equation~(\ref{Eq:D_t0}), we have
\begin{equation}\label{Eq:Spectral Evolution}
\begin{array}{*{20}{c}}
{{\rm{Case\;(I)}}:}&{\beta(\tilde{t}_{\rm obs})  = \left\{ {\begin{array}{*{20}{c}}
{a_1+7.9\tilde{\kappa}{\tilde{t}_{\rm obs}}/{\tilde{t}_{c,r}} ,}&{10{\rm{keV}} \leqslant( {\hat \alpha  - \hat \beta })\tilde{E}_{0,r}(\tilde{t}_{\rm obs}),}\\
{a_2 + \tilde{b}{\tilde{t}_{{\rm{obs}}}}/{\tilde{t}_{c,r}} + c\ln ( {1 + {\tilde{t}_{{\rm{obs}}}}/\tilde{t}_{c,r}}  ),}&{0.3{\rm{keV}}< (\hat \alpha  - \hat \beta )\tilde{E}_{0,r}(\tilde{t}_{\rm obs}) < 10{\rm{keV}},}\\
{-  \hat \beta  - 1},&{0.3{\rm{keV}} \geqslant ( {\hat \alpha  - \hat \beta } )\tilde{E}_{0,r}(\tilde{t}_{\rm obs}),}
\end{array}} \right.}\\
{{\rm{Case\;(II)}}:}&{\beta(\tilde{t}_{\rm obs}) = a_1+7.9\tilde{\kappa}{\tilde{t}_{{\rm{obs}}}}/{\tilde{t}_{c,r}} ,}
\end{array}
\end{equation}
where
\begin{equation}
\tilde{E}_{0,r}(\tilde{t}_{\rm obs})=\frac{1}{1+z}\frac{E'_0(r)D_{t_0}}{1+\tilde{t}_{\rm obs}/\tilde{t}_{c,r}}=\frac{\tilde{E}_{0,r}(\tilde{t}_{\rm obs}=0)}{1+\tilde{t}_{\rm obs}/\tilde{t}_{c,r}},
\end{equation}
\begin{equation}
\tilde{\kappa}  = \frac{0.3{\rm keV}}{\tilde{E}_{0,r}(\tilde{t}_{\rm obs}=0)},
\end{equation}
$\tilde{b} =- 0.4706\tilde{\kappa}$,
and $\tilde{E}_{0,r}(\tilde{t}_{\rm obs}=0)$ is the observed characteristic photon energy
of the radiation spectrum at $\tilde{t}_{\rm obs}=0$.
With $\tilde{t}_{c,r}=t_{c,r}+t_0-t_{{\rm obs}, r}$, $D_{t_0}={2\Gamma}/[{1+(t_0-t_{{\rm obs}, r})/t_{c,r}}]$,
and $t_0\geqslant t_{{\rm obs}, r}$,
Equation~(\ref{Eq:Spectral Evolution}) is applicable to describe the spectral evolution
for the radiation from an EFCS located at any $r$,
where
\begin{equation}
t_{{\rm obs}, r}\equiv (1 + z)\int_{{r_0}}^r {[1-\beta_{\rm jet}(l)]}\frac{dl}{c\beta_{\rm jet}(l)}
\end{equation}
is the observed time for the first photon from a radiating jet shell located at $r$.
If $t_0=t_{{\rm obs}, r}$, we have $a_1=- \hat \alpha- 1 + 7.9\tilde{\kappa}$
and $a_2=a$ with $\kappa$ being replaced by $\tilde{\kappa}$ for an EFCS based on Equation~(\ref{Eq:Spectral Evolution_ini}).

In general, the shell may radiate from $r_0$ to $r_e$ with $r_e>r_0$.
An expanding jet in our work is modelled with a series of jet shells
located at radius $r_0,\;r_1=r_0+\beta_{\rm jet}(r)c{\delta t},\;r_2=r_1+\beta_{\rm jet}(r_1)c{\delta t},\;\cdot\cdot\cdot,r_n=r_{n-1}+\beta_{\rm jet}(r_{n-1})c{\delta t},\;\cdot\cdot\cdot$
with appearing time $t=0{\rm s},\;{\delta t},\;{2\delta t},\;\cdot\cdot\cdot, n{\delta t},\;\cdot\cdot\cdot $, respectively.
The radiation behavior of the jet shell during the time interval $[(n-1){\delta t}, n{\delta t}]$ does not change.
This behavior is similar to that of an EFCS's radiation discussed above.
Then, the radiation of our jet can be regarded
as the radiation from a series of EFCSs
located at $r_0,\;r_1,\;r_2,\;\cdot\cdot\cdot,r_n,\;\cdot\cdot\cdot$ with appearing time
$t=0{\rm s},\;{\delta t},\;{2\delta t},\;\cdot\cdot\cdot, n{\delta t},\;\cdot\cdot\cdot $.
Thus, we would like to use the observed photon energy $E_0(\tilde{t}_{\rm obs})$ of the radiation spectrum
to replace $\tilde{E}_{0,r}(\tilde{t}_{\rm obs})$, i.e.,\\
Case (I):
\begin{equation}\label{Eq:Spectral Evolution_reality}
\beta(\tilde{t}_{\rm obs})  = \left\{ {\begin{array}{*{20}{c}}
{a_1+7.9\hat{\kappa}{\tilde{t}_{{\rm{obs}}}}/{\tilde{t}_{c}} ,}&{10{\rm{keV}} \leqslant( {\hat \alpha  - \hat \beta })fE_0(\tilde{t}_{\rm obs}),}\\
{a_2 + \hat{b}{\tilde{t}_{{\rm{obs}}}}/{\tilde{t}_{c}} + c\ln ( {1 + {\tilde{t}_{{\rm{obs}}}}/\tilde{t}_{c}}  ),}&{0.3{\rm{keV}}< (\hat \alpha  - \hat \beta )fE_0(\tilde{t}_{\rm obs}) < 10{\rm{keV}},}\\
{-  \hat \beta  - 1},&{0.3{\rm{keV}} \geqslant ( {\hat \alpha  - \hat \beta } )fE_0(\tilde{t}_{\rm obs}),}
\end{array}} \right.
\end{equation}
Case (II):
\begin{equation}\label{Eq:Spectral Evolution_II}
\beta(\tilde{t}_{\rm obs}) = a_1+7.9\hat{\kappa}{\tilde{t}_{{\rm{obs}}}}/{\tilde{t}_{c}},
\end{equation}
where $a_1$ and $a_2$ are constants, $\tilde{t}_c$ is the decay timescale of the phase with $\tilde{t}_{{\rm{obs}}}\geqslant 0$,
$E_0(\tilde{t}_{\rm obs})=E_{0,0}/(1+\tilde{t}_{\rm obs}/\tilde{t}_c)$ with $E_{0,0}=E_0(\tilde{t}_{\rm obs}=0)$ being the observed photon energy at $\tilde{t}_{\rm obs}=0$, $\hat{\kappa}  = {0.3{\rm keV}}/(fE_{0,0})$,
and $\hat{b} =- 0.4706\hat{\kappa}$.
For situations with Case (I), the value of $( {\hat \alpha  - \hat \beta })E_0$ is the observed break energy of Band function;
for situations with Case (II), the value of $E_0$ is the observed cutoff energy of CPL spectrum.
The value of $f\sim 1$ is introduced by considering that the observed flux is
from a series of EFCSs located at different $r$ with different $t_{{\rm obs},r_n}$.
As discussed in Section~\ref{Sec:Testing},
the exact value of $f$ depends on the behavior of jet's dynamics and radiation,
and thus is difficult to estimate in reality.
It is interesting to note that the value of $f$ is around unity for our studying cases (see Figure~3).
Moreover, Equation~(\ref{Eq:Spectral Evolution_reality}) with $f=1$ presents a well estimation about the spectral evolution (see Figure~3). Then, we suggest to use $f=1$ in practice.

\emph{Equations~(\ref{Eq:Spectral Evolution_reality}) and (\ref{Eq:Spectral Evolution_II})
are our obtained analytical formula of the spectral evolution in the steep decay phase.
}Since Equation~(\ref{Eq:Spectral Evolution_II}) is involved in Equation~(\ref{Eq:Spectral Evolution_reality}),
we only test Equation~(\ref{Eq:Spectral Evolution_reality}) with our numerical simulations.
For Equation~(\ref{Eq:Spectral Evolution_reality}),
if $10{\rm{keV}} \leqslant( {\hat \alpha  - \hat \beta })fE_{0,0}$ is satisfied,
the value of $a_1$ would be the spectral index at $\tilde{t}_{\rm obs}=0$.
In this situation, $a_1=\beta(\tilde{t}_{\rm obs}=0)$ is adopted in our testing process
and
the value of $a_2$ is appropriately took in order to
remain the continuity of $\beta(\tilde{t}_{\rm obs})$ at $fE_0=10{\rm{keV}}/( {\hat \alpha  - \hat \beta })$.
If $0.3{\rm{keV}}< (\hat \alpha  - \hat \beta )fE_{0,0} < 10{\rm{keV}}$ is satisfied,
the value of $a_2$ would be the spectral index at $\tilde{t}_{\rm obs}=0$.
In this situation, $a_2=\beta(\tilde{t}_{\rm obs}=0)$ is adopted in our testing process.
It is interesting to find that for significantly large value of $E_{0,0}$, the value of $\hat{\kappa}$ would be low and thus $\beta(\tilde{t}_{\rm obs})\approx a_2 + c\ln ( {1 + {\tilde{t}_{{\rm{obs}}}}/\tilde{t}_{c}}  )$ can be found.

\section{Testing}\label{Sec:Testing}

In this section, we test Equation~(\ref{Eq:Spectral Evolution_reality}) based on the numerical simulations.
Figure~2 shows the evolution of $\beta$ for an EFCS.
Here, $t_{\rm obs}=0$ is the beginning of the steep decay phase dominated by the shell curvature effect.
For each part of this figure,
the upper panel plots the integrated flux in $0.3-10{\rm keV}$ energy band
and the lower panel shows the spectral index $\beta$.
In the left part, the violet ``$\Box$'', red ``$\circ$'', black ``$+$'', and green ``$\times$'' represent the data from
the numerical simulations with $2E'_{0,0}\Gamma_0/(1+z)=0.1\rm keV$, $1\rm keV$, $5\rm keV$, and $50\rm keV$, respectively.
The violet, red, black, and green solid lines represent the value of $\beta$ estimated
with Equation~(\ref{Eq:Spectral Evolution_reality}), $\tilde{t}_c=t_{c,r_0}$, $f=1$,
and $E_{0,0}=0.1\rm keV$, $1\rm keV$, $5\rm keV$, and $50\rm keV$, respectively.
It can be found that Equation~(\ref{Eq:Spectral Evolution_reality})
can present a well estimation about the spectral evolution for an EFCS.
In the right part of this figure, we plot the light curves and spectral evolution with $t_0=t_{c, r_0}$.
The meaning of symbols are the same as those in the left part.
In the lower panel of right part, the decay timescale is $\tilde{t}_{c}=t_{c, r_0}+t_0=2t_{c,r_0}$ according to Equation~(\ref{Eq:D_t0}).
In addition, one can find $E_{0,0}=E'_{0,0}D_{t_0}/(1+z)=E'_{0,0}\Gamma_0/(1+z)$ at $t_{\rm obs}=t_0$.
Then, we plot the value of $\beta$ estimated with Equation~(\ref{Eq:Spectral Evolution_reality}), $\tilde{t}_{c}=2t_{c, r_0}$,
$f=1$, and $E_{0,0}=0.05\rm keV$ (violet solid line), $0.5\rm keV$ (violet solid line),
$2.5\rm keV$ (black solid line), and $25\rm keV$ (red solid line), respectively.
It can be found that Equation~(\ref{Eq:Spectral Evolution_reality}) presents a well estimation
about the spectral evolution in the steep decay phase for an EFCS.

In Figure~3, we show the results for situations with a spherical thin shell radiating from $r_0$ to $2r_0$,
where the data from numerical simulations with different $s$ and $w$ are plotted in sub-figures (a)-(i), respectively.
In each sub-figure,
the upper-left panel shows the evolution of integrated flux in $0.3-10{\rm keV}$ energy band,
and the lower-left panel (right part) shows the spectral evolution for $t_{\rm obs}\geqslant 0$ ($t_{\rm obs}\geqslant t_p$).
Here, $t_p$ is the peak time of the integrated flux in $0.3-10{\rm keV}$ energy band,
and $t_p=2.32t_{c,r}$, $1.01t_{c,r}$, and $0.50t_{c,r}$ are found
for situations with $s=-1$, $0$, and $1$, respectively.
It should be noted that the phase with $t_{\rm obs}\geqslant t_p$ is dominated by the shell curvature effect in our simulations
(\citealp{Uhm2015}; \citealp{Lin_DB-2017-Mu_HJ}).
The violet ``$\Box$'', red ``$\circ$'', black ``$+$'', and green ``$\times$'' in Figure~3 represent the data from
the numerical simulations with $2E'_{0,0}\Gamma_0/(1+z)=0.1\rm keV$, $1\rm keV$, $5\rm keV$, and $50\rm keV$, respectively.
For comparison, the $\beta$ estimated with Equation~(\ref{Eq:Spectral Evolution_reality}) and $E_{0,0}=2E'_0(2r_0)\Gamma(2r_0)$
is shown with solid lines in the right part of each sub-figures,
where $\tilde{t}_c=6.41t_{c,r_0}$, $1.96t_{c,r_0}$, and $0.69t_{c,r_0}$
are adopted for situations with $s=-1$, $0$, and $1$ (\citealp{Lin_DB-2017-Mu_HJ}), respectively.
In general, Equation~(\ref{Eq:Spectral Evolution_reality}) with $f\sim 1$
presents a well estimation about the spectral evolution according to the results showed in Figure~3.
Then, we can conclude that Equation~(\ref{Eq:Spectral Evolution_reality}) can describe the spectral evolution.

The values of $f$ adopted in Figure~3 are estimated based on the following discussion.
For the phase with $t_{\rm obs}\geqslant t_p$, we have $t_0=t_p$ and $\tilde{t}_{\rm obs}=t_{\rm obs}-t_p$.
As discussed in Section~\ref{Sec:Evolution of Spectra},
the observed flux in the phase with $\tilde{t}_{\rm obs}\geqslant 0$
is from a series of EFCSs located at
$r_0,\;r_1,\;r_2,\;\cdot\cdot\cdot,r_n,\;\cdot\cdot\cdot$
with appearing time $t=0{\rm s},\;{\delta t},\;{2\delta t},\;\cdot\cdot\cdot, n{\delta t},\;\cdot\cdot\cdot $.
Then, any EFCS can exert more or less influence on the spectral evolution in the steep decay phase.
It should be noted that the observed time for the first photon from an EFCS located at $r$ is $t_{{\rm obs}, r}$.
Thus, the spectral evolution for the radiation from an EFCS located at $r$
can be described with Equation~(\ref{Eq:Spectral Evolution}) by adopting $\tilde{t}_{c,r}=t_{c,r}+t_0-t_{{\rm obs}, r}$, $D_{t_0}={2\Gamma}/[{1+(t_0-t_{{\rm obs}, r})/t_{c,r}}]$, and $t_0\geqslant t_{{\rm obs}, r}$.
That is to say, the spectral evolution in the phase with $\tilde{t}_{\rm obs}\geqslant 0$
for the radiation from an EFCS is controlled by two parameters:
$\tilde{E}_{0,r}(\tilde{t}_{\rm obs}=0)=2\Gamma(r) \hat E'_0(r)/[1+(t_p-t_{{\rm obs}, r})/t_{c,r}]$ and $\tilde{t}_{c,r}=t_{c,r}+t_p-t_{{\rm obs}, r}$.
Since the radiation of jet shell can be regarded as the radiation from
a series of EFCSs located at different $r$ with different $t_{{\rm obs},r}$,
different pattern of $r$-dependent $\tilde{E}_{0,r}(\tilde{t}_{\rm obs}=0)$ and $\tilde{t}_{c,r}$ for EFCSs
would require a different value of $f$ to describe the spectral evolution as Equation~(\ref{Eq:Spectral Evolution_reality}).
Taking the situation with $s=0$ and $w=0$ as an example,
it can be found that the decay timescale $\tilde{t}_{c,r}$ for radiation from any EFCS is the same, i.e., $\tilde{t}_{c,r}=\tilde{t}_{c,r_e}$.
However, the value of $\tilde{E}_{0,r}(\tilde{t}_{\rm obs}=0)=2\Gamma_0E'_{0,0}t_{c,r}/\tilde{t}_{c,r}\propto r$ increases with the location of EFCSs.
Then, the spectral evolution should be fitted with Equation~(\ref{Eq:Spectral Evolution_reality}) and $f\lesssim1$
if $E_{0,0}=2E'_0(2r_0)\Gamma(2r_0)$ is satisfied, where $E_{0,0}=2E'_0(2r_0)\Gamma(2r_0)$ is found in our simulation.
This behavior can be found in Figure~3a,
where Equation~(\ref{Eq:Spectral Evolution_reality}) with $f=0.85$ presents a better estimation about the spectral evolution.
For the situation with $s=0$ and $w=-1$,
$\tilde{t}_{c,r}$ and $\tilde{E}_{0,r}(\tilde{t}_{\rm obs}=0)=2\Gamma_0E'_{0,0}rt_{c,r}/(\tilde{t}_{c,r}r)$
remain constants for different EFCSs.
Then, the spectral evolution can be described with Equation~(\ref{Eq:Spectral Evolution_reality}),
$E_{0,0}=2E'_0(2r_0)\Gamma(2r_0)$, and $f=1$.
This can be found in Figure~3b,
where Equation~(\ref{Eq:Spectral Evolution_reality}) with $f=1$ presents a perfect estimation about the spectral evolution.
Then, Equation~(\ref{Eq:Spectral Evolution_reality}) with $f\neq 1$ is not shown in this sub-figure,
and the bottom-right panel of Figure~3b is empty.
The value of $f$ adopted in other situations can be analyzed in the same way as that shown above.
In reality, however, the exact value of $f$ is difficult to estimate.
Figure~3 shows that Equation~(\ref{Eq:Spectral Evolution_reality}) with $f=1$ can present a well estimation about the spectral evolution. Then, we conclude that Equation~(\ref{Eq:Spectral Evolution_reality}) with $f=1$ can present a well estimation about the spectral evolution in the steep decay phase.

\section{Conclusions and Discussion}\label{Sec:conclusion}
This work focuses on the spectral evolution in the steep decay phase shaped by the curvature effect.
We study the radiation from a spherical relativistic thin shell with significantly large jet opening angle ($\theta_{\rm jet}>>1/\Gamma$). In addition, we assume that (1) the central axis of jet shell coincides with the observer's line of sight;
(2) the jet shell has no $\theta$-dependent spectral parameters and Lorentz factor.
For shells with a cutoff power-law intrinsic radiation spectrum,
we find that the spectral evolution can be described as Equation~(\ref{Eq:Spectral Evolution_II}), i.e.,
$\beta(\tilde{t}_{\rm obs})\propto 7.9\hat{\kappa}{\tilde{t}_{{\rm{obs}}}}/{\tilde{t}_{c}}$.
This equation reveals that $\beta(\tilde{t}_{{\rm{obs}}})$ is a linear function of the observer time.
For shells with Band function intrinsic radiation spectrum,
the spectral evolution is complex
and can be described as Equation~(\ref{Eq:Spectral Evolution_reality}) with $f=1$, i.e.,
(1) for $10{\rm{keV}} \leqslant( {\hat \alpha  - \hat \beta })E_0(\tilde{t}_{{\rm{obs}}})$,
$\beta(\tilde{t}_{\rm obs})\propto 7.9\hat{\kappa}{\tilde{t}_{{\rm{obs}}}}/{\tilde{t}_{c}}$;
(2) for $0.3{\rm{keV}}< ( {\hat \alpha  - \hat \beta })E_0(\tilde{t}_{{\rm{obs}}}) < 10{\rm{keV}}$,
$\beta(\tilde{t}_{\rm obs})\propto - 0.47\hat{\kappa}{\tilde{t}_{{\rm{obs}}}}/{\tilde{t}_{c}} + 0.35( {\hat \alpha  - \hat \beta })\ln ( {1 + {\tilde{t}_{{\rm{obs}}}}/\tilde{t}_{c}}  )$ or $\beta(\tilde{t}_{\rm obs})\propto \ln ( {1 + {\tilde{t}_{{\rm{obs}}}}/\tilde{t}_{c}}  )$ for significantly large $E_0(\tilde{t}_{{\rm{obs}}}=0)$;
(3) for $0.3{\rm{keV}} \geqslant ( {\hat \alpha  - \hat \beta })E_0(\tilde{t}_{{\rm{obs}}})$, $\beta(\tilde{t}_{\rm obs})=- \hat \beta  - 1$.
The spectral evolution in this situation depends on the break energy $( {\hat \alpha  - \hat \beta })E_0$
in the observed Band function spectrum.
The above results are tested by the data from our numerical simulations.

Our formula can be used to confront the curvature effect with observations and
estimate the decay timescale of the steep decay phase.
In observations, the steep decay phase has been observed in the decay phase of prompt emission phase and flares in GRBs.
Since the value of $E_0$ ($\sim 0.3-1\rm MeV$) in the prompt emission phase is significantly larger than $10\rm keV$,
the $\beta$ in the steep decay phase of prompt emission would be a linear function of observer time based on Equation~(\ref{Eq:Spectral Evolution_reality}).
This behavior has been observed in a number of bursts,
such as GRBs 050814 (\citealp{Zhang2009}, please see the spectral evolution in the $\beta-t_{\rm obs}$ space), 051001 etc.
The linear relation between $\beta$ and $\tilde{t}_{\rm obs}$ is also found in the steep decay of flares (e.g., \citealp{Mu2016}).
For a linear function $\beta$ of $\tilde{t}_{\rm obs}$, one can obtain the slope of the linear function,
which equals to $7.9\hat{\kappa}/{\tilde{t}_{c}}={2.37{\rm keV}}/[{{E}_{0}(\tilde{t}_{\rm obs}=0)}{\tilde{t}_{c}}]$ based on Equation~(\ref{Eq:Spectral Evolution_reality}) or (\ref{Eq:Spectral Evolution_II}).
Then, the decay timescale $\tilde{t}_{c}$ of our studying phase can be estimated
if the value of ${E}_{0}(\tilde{t}_{\rm obs}=0)$ is known.
We fit the spectral evolution in the decay phase of a flare ($\sim 172\rm s$) in GRB~060904B (\citealp{Mu2016}) with a linear function.
The fitting result, i.e., $\beta=0.84+0.020\tilde{t}_{\rm obs}$ with $\tilde{t}_{\rm obs}=t_{\rm obs}-t_p$, is shown in the left panel of Figure~4 with red solid line.
Then, we have ${{E}_{0}(\tilde{t}_{\rm obs}=0)}{\tilde{t}_{c}}=119\,{\rm keV\cdot s}$, which is around that found in \cite{Mu2016}, i.e., $2.54{\rm keV}\times 65.39{\rm s}$.
For a flare ($\sim 116\rm s$) in GRB 131030A,
the spectrum at the beginning of the steep decay phase can be fitted with a Band function.
With the fitting result found in \cite{Mu2016}, i.e., $\hat{\alpha}=-0.91$, $\hat{\beta}=-3.19$, and ${E}_{0}(\tilde{t}_{\rm obs}=0)=1.95$,
we plot the evolution of $\beta$ based on Equation~(\ref{Eq:Spectral Evolution_reality}) in the right panel of Figure~4 with red solid line. Here, we adopt $\tilde{t}_{c}=54\rm s$ (\citealp{Mu2016}),
which can also be roughly estimated based on the flux evolution.
It can be found that Equation~(\ref{Eq:Spectral Evolution_reality})
describes the spectral evolution approximately,
which may reveal that more appropriate value of parameters (e.g., $\tilde{t}_{c}$) may be required for this source.
The agreement of our analytical formula and observational data
shows that the assumption (2) given at the beginning of Section~2 (i.e., the jet shell has no $\theta$-dependent spectral parameters and Lorentz factor) is applicable in reality.

\acknowledgments
We thank the anonymous referee of this work for beneficial suggestions that improved the paper.
This work is supported by the National Basic Research Program of China (973 Program, grant No. 2014CB845800),
the National Natural Science Foundation of China (Grant Nos. 11403005, 11533003, 11673006, 11573023, 11473022, U1331202),
the Guangxi Science Foundation (Grant Nos. 2016GXNSFDA380027, 2014GXNSFBA118004, 2016GXNSFFA380006, 2014GXNSFBA118009, 2013GXNSFFA019001), and the Innovation Team and Outstanding Scholar Program in Guangxi Colleges.



\clearpage
\begin{figure}\label{Fig:Compare_beta}
\plotone{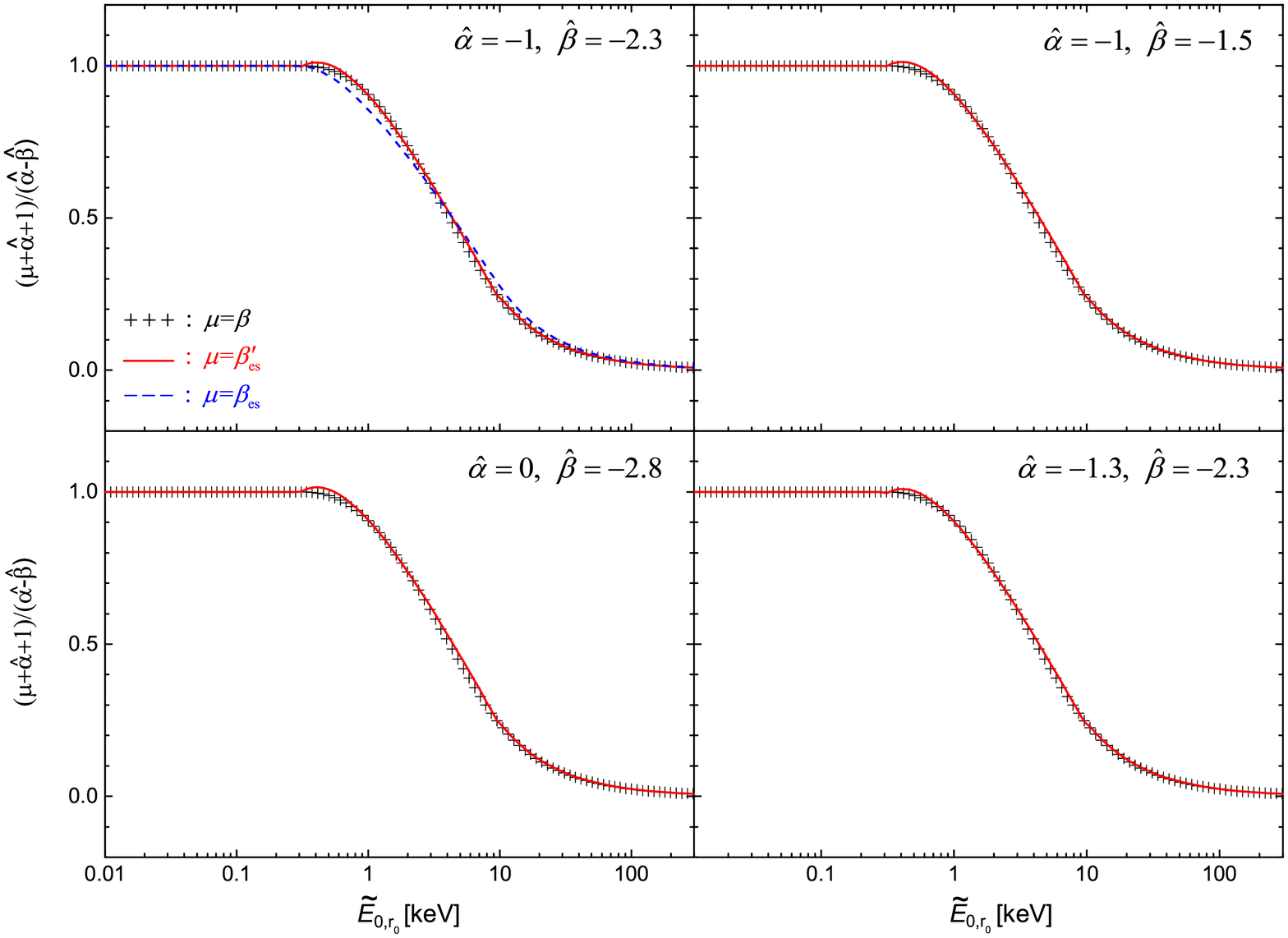}
\caption{Comparison of $\beta$, $\beta_{\rm es}$, and $\beta'_{\rm es}$ for different $\hat\alpha$ and $\hat\beta$ in the Band function spectrum.
}
\end{figure}

\clearpage
\begin{figure}\label{Fig:Thin_Shell_Case}
\includegraphics[angle=0,scale=0.350,width=0.5\textwidth,height=0.40\textheight]{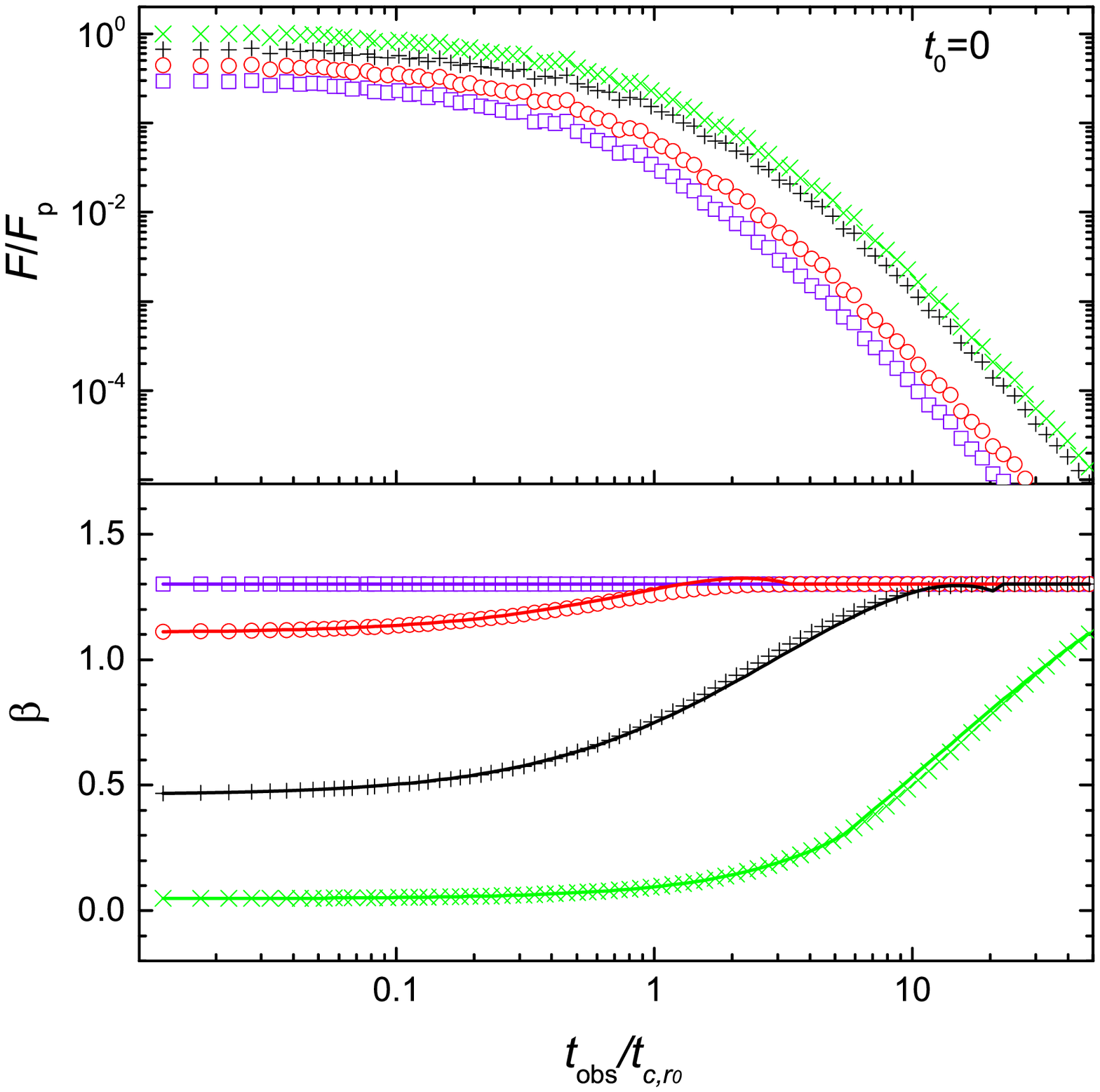}
\includegraphics[angle=0,scale=0.350,width=0.5\textwidth,height=0.40\textheight]{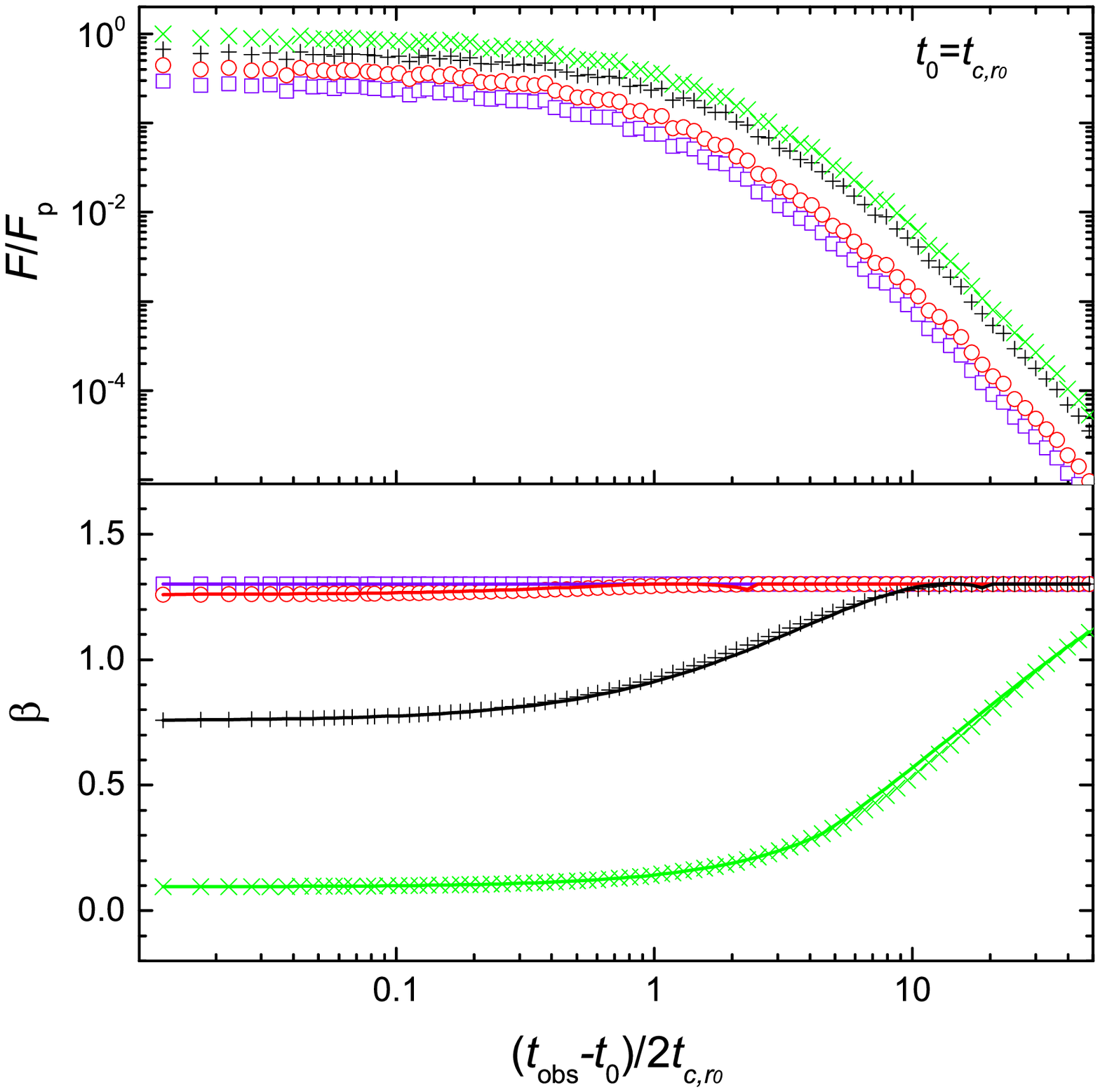}
\caption{Comparison of $\beta$ from simulations and that estimated with Equation~(36)
for an EFCS, where $f=1$ is took. The data with $t_0=0$ and $t_0=t_{c,r_0}$ are shown in the left and right panels, respectively.
For clarity, the flux is shifted by dividing $1.5$ for adjacent light curves in the plot.
}
\end{figure}

\clearpage
\begin{figure}\label{Fig:Thin_Shell_Case}
\plotone{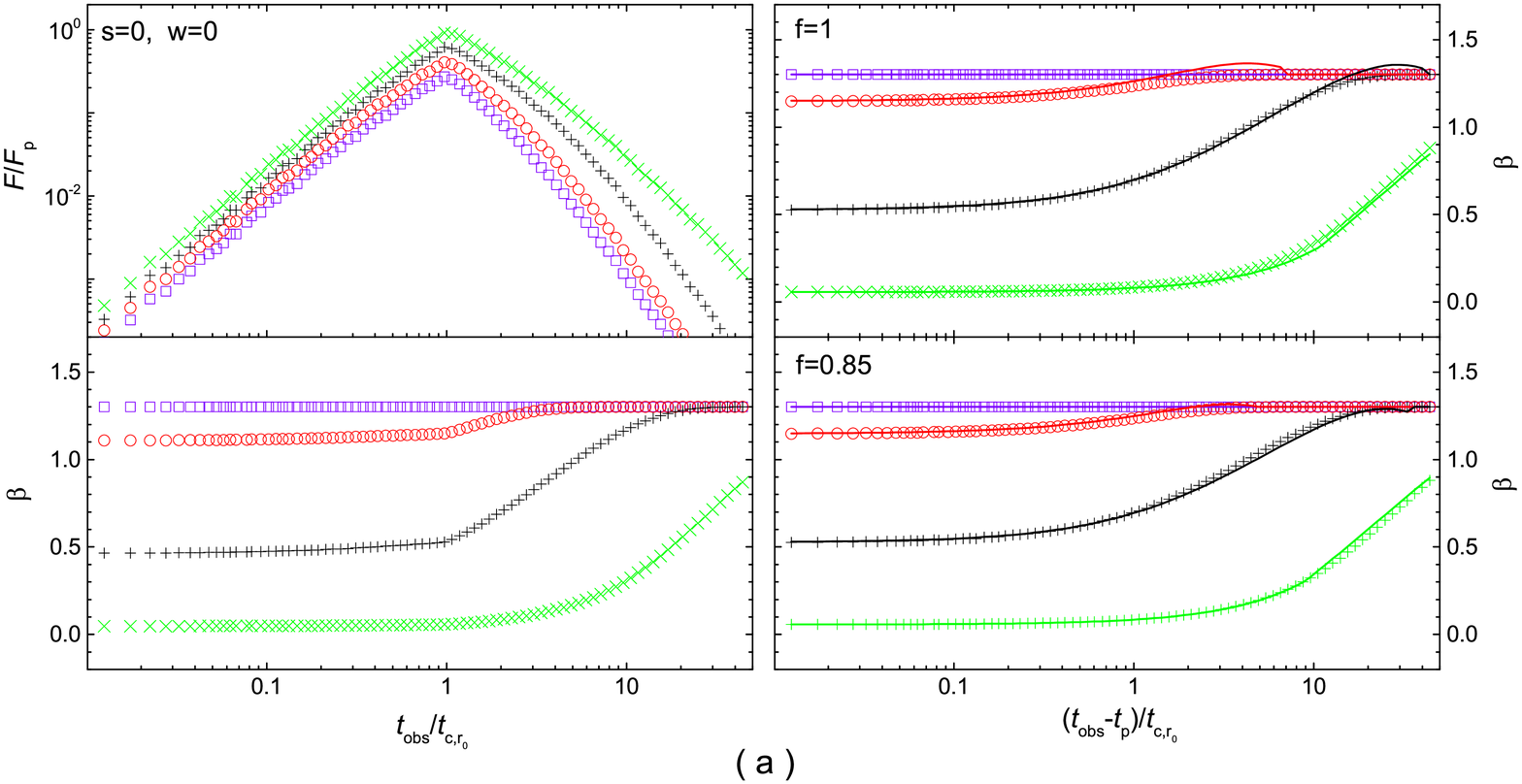}
\plotone{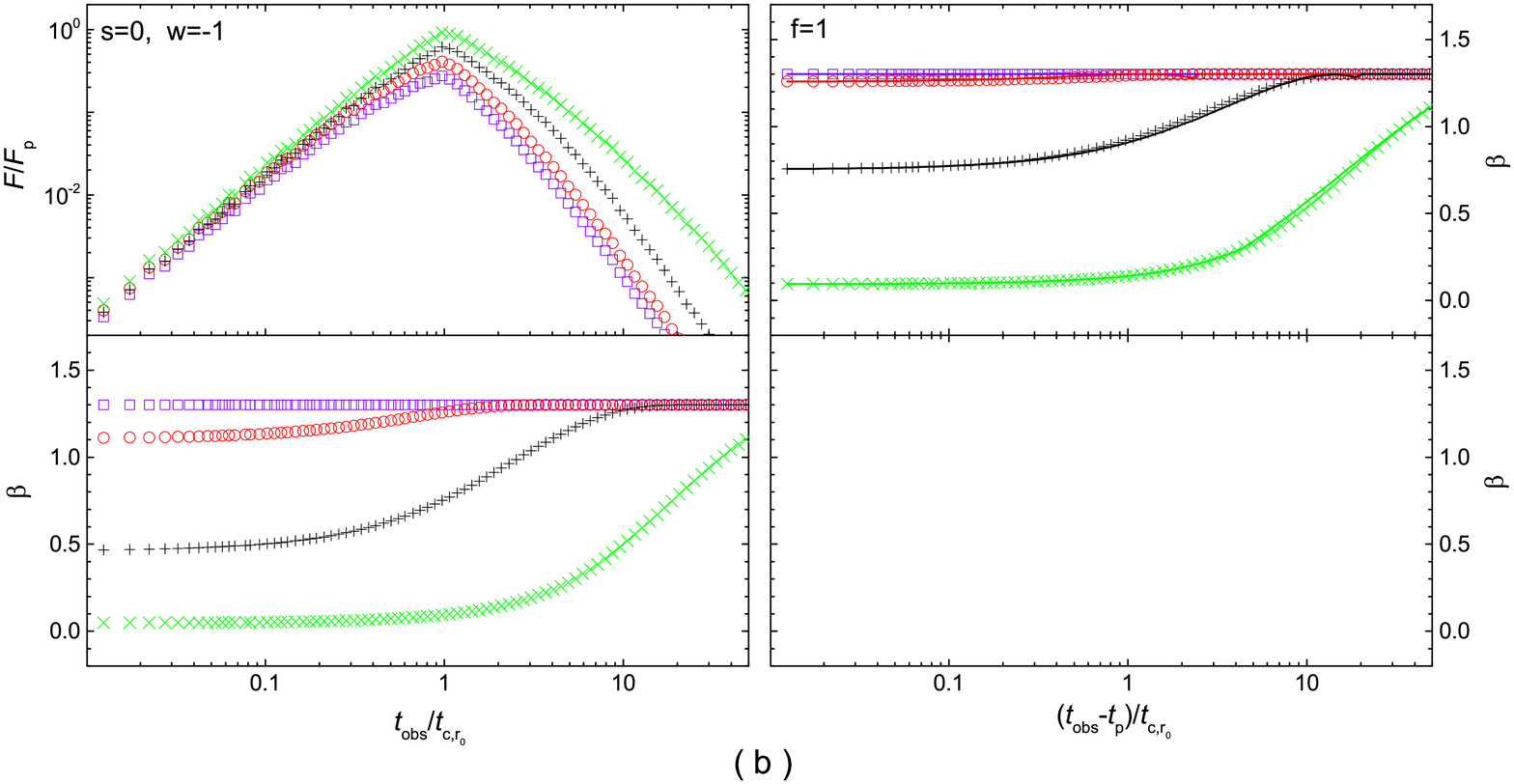}
\caption{\emph{Left Panel}: Evolution of flux and spectral index for a thin jet shell radiating from $r_0$ to $2r_0$;
\emph{Right Panel}: Comparison of $\beta$ from simulations and that estimated with Equation~(36) for the steep decay phase,
where different value of $s$ and $w$ in the simulations are studied.
For clarity, the flux is shifted by dividing $1.5$ for adjacent light curves in the plot.
}
\end{figure}
 \addtocounter{figure}{-1}
\begin{figure}
\plotone{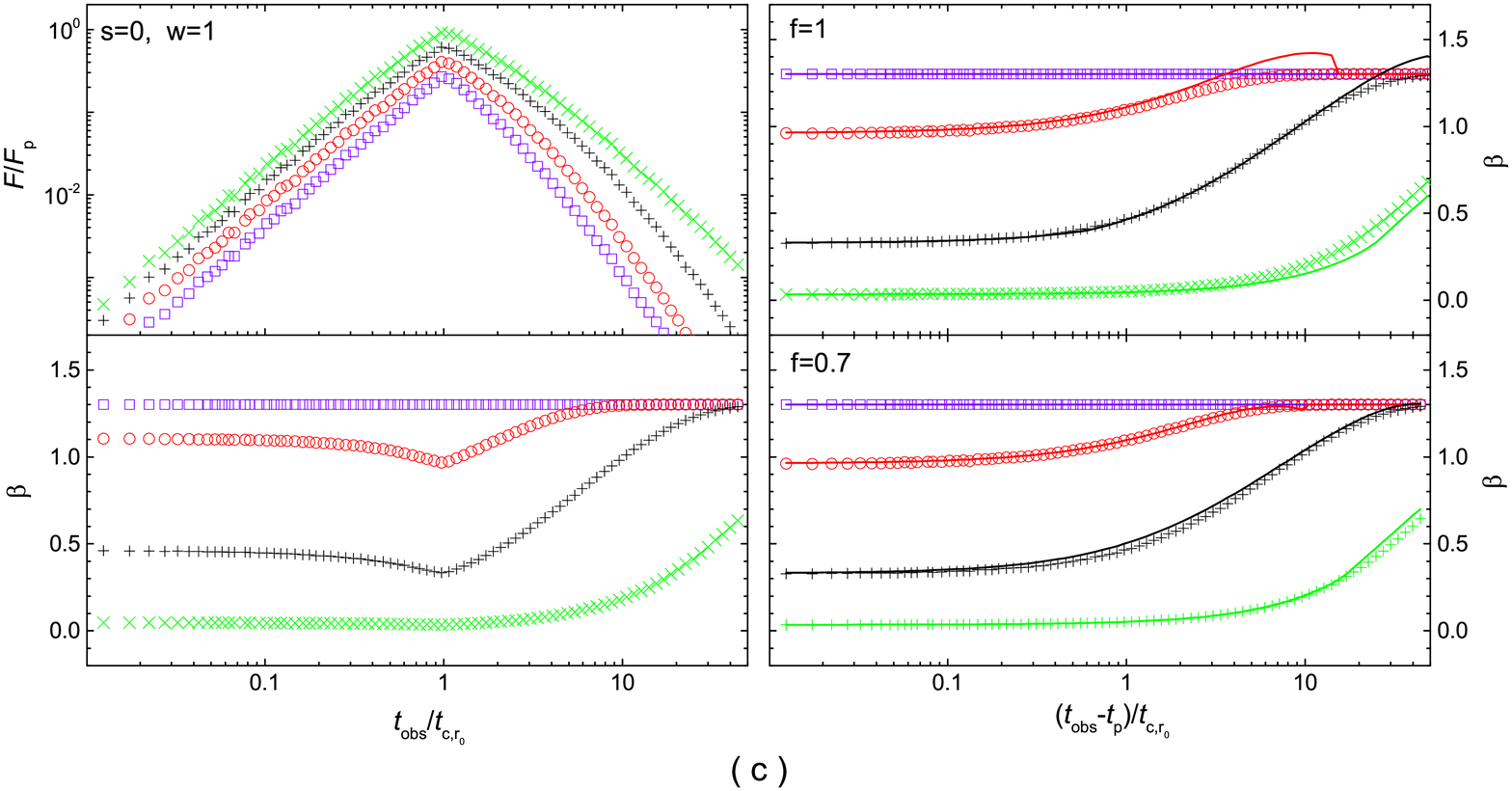}
\plotone{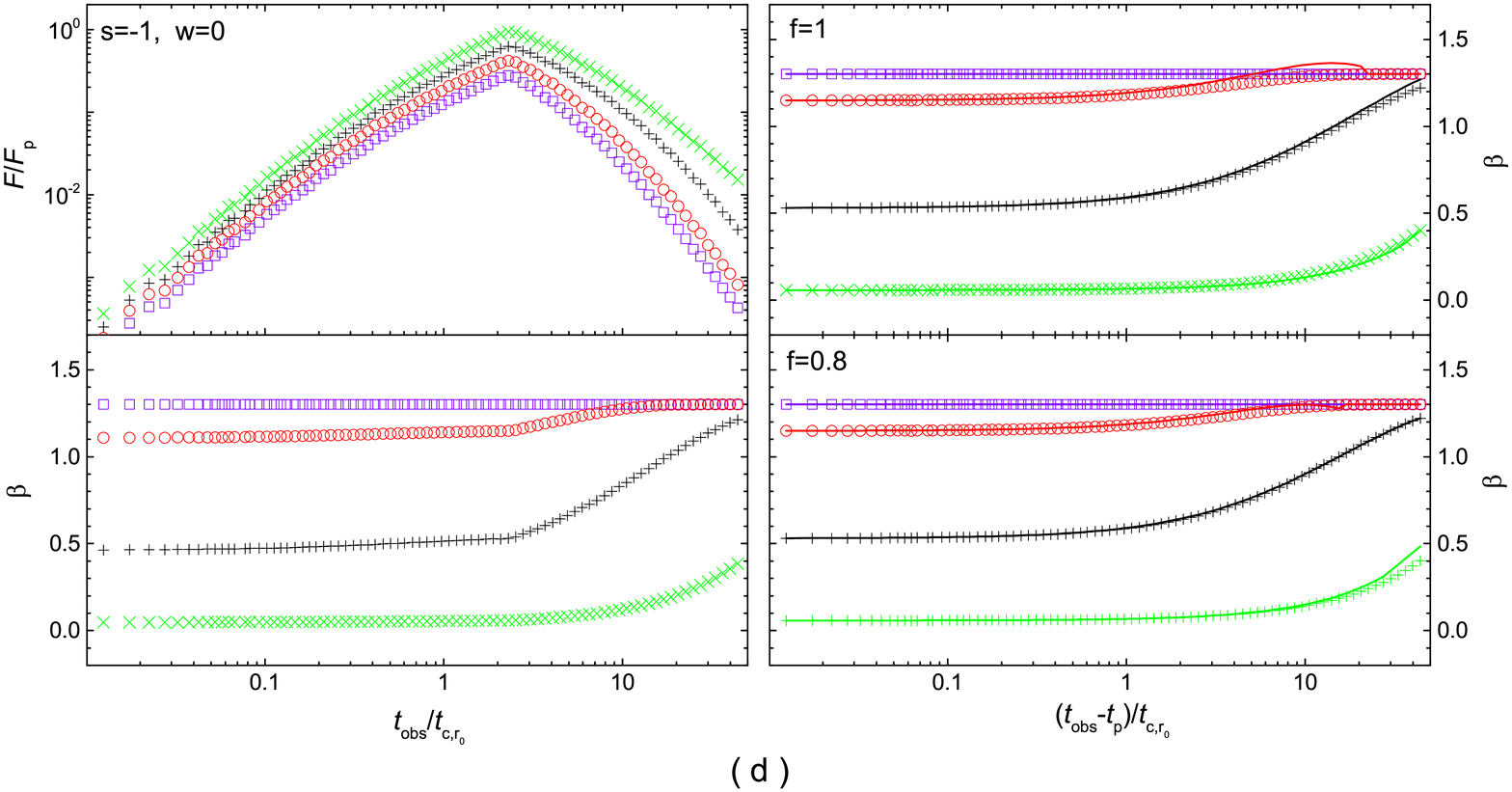}
\caption{\emph{Continued}
}
\end{figure}
\addtocounter{figure}{-1}
\begin{figure}
\plotone{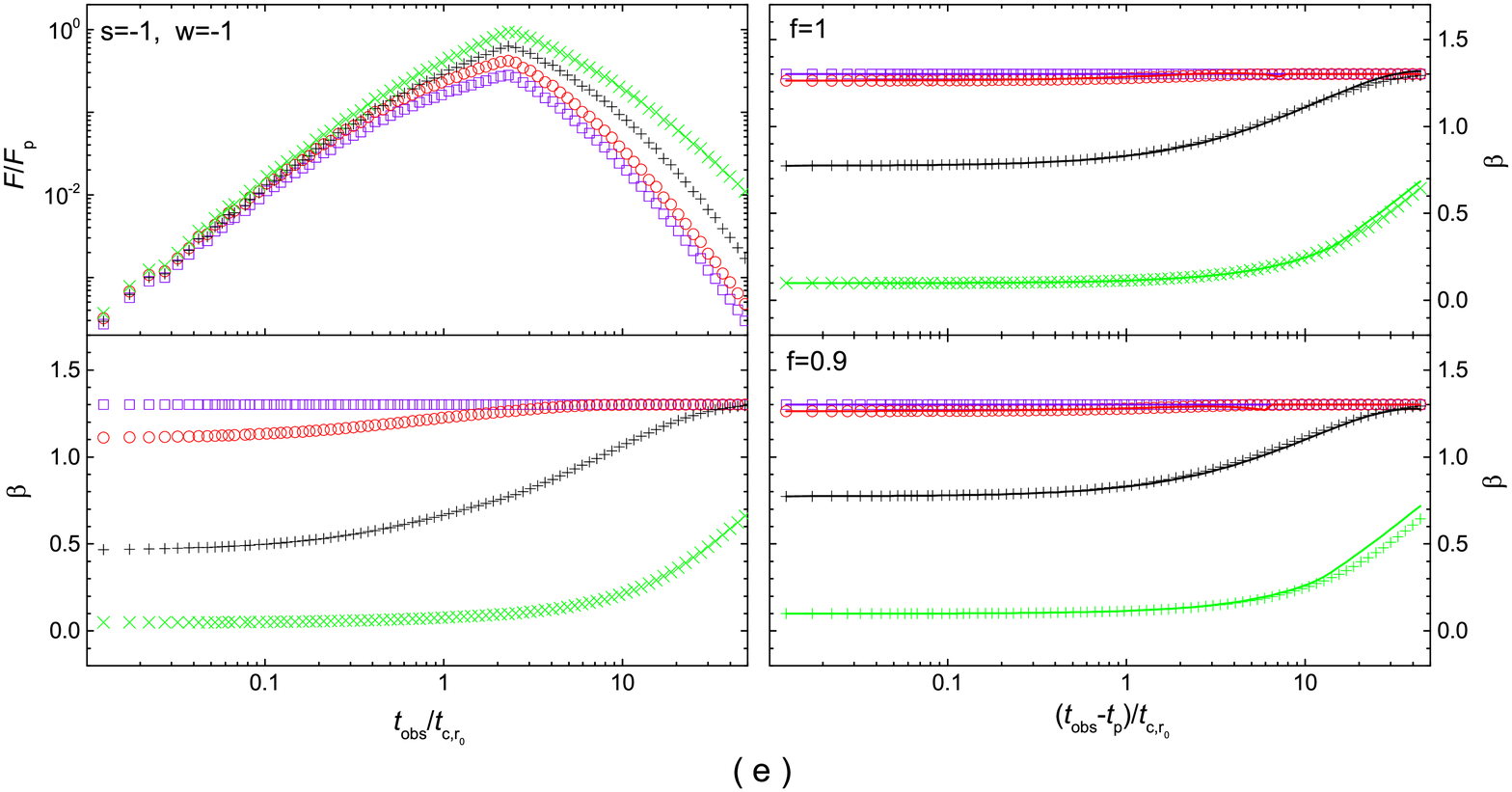}
\plotone{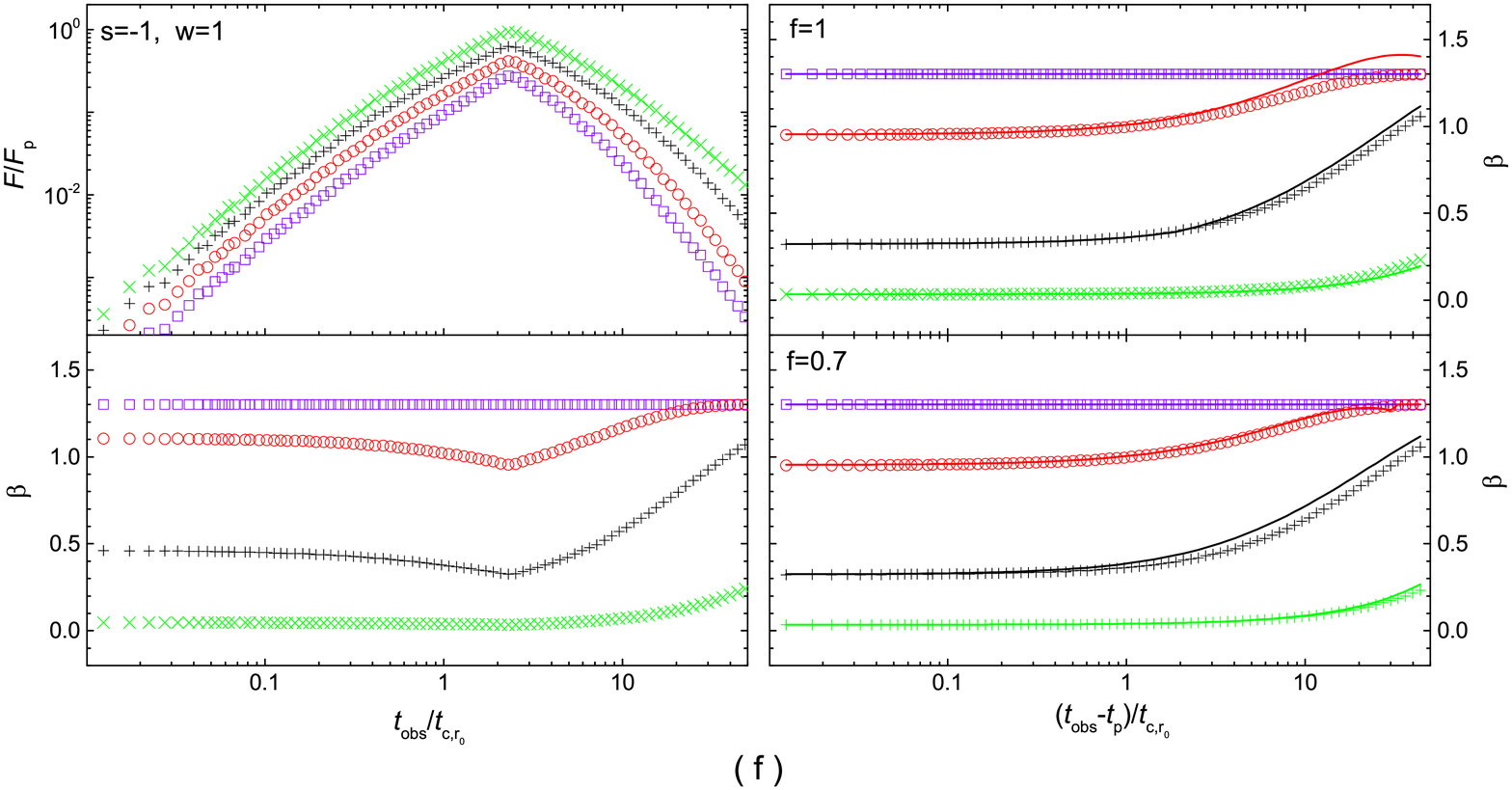}
\caption{\emph{Continued}
}
\end{figure}
 \addtocounter{figure}{-1}
\begin{figure}
\plotone{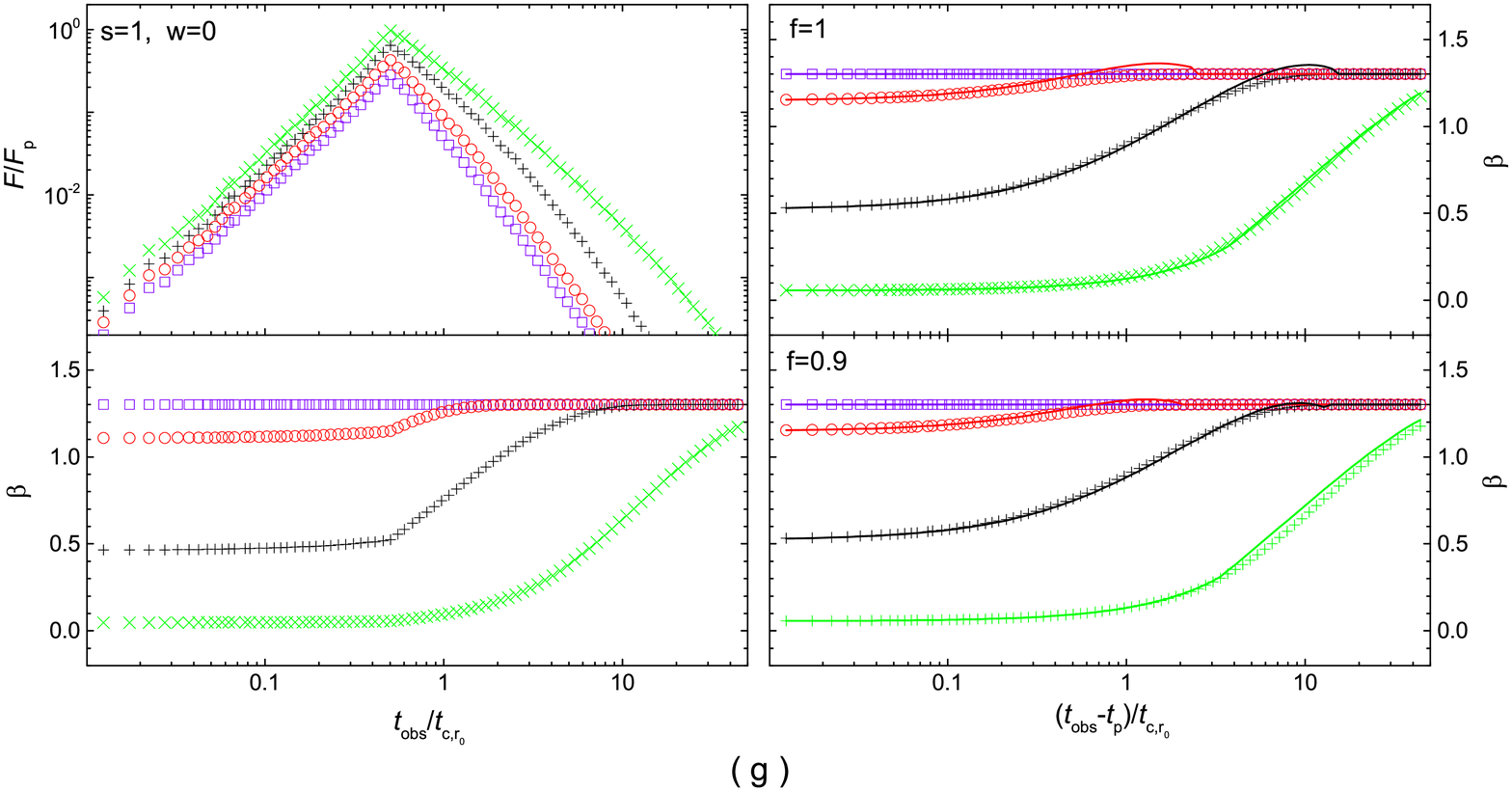}
\plotone{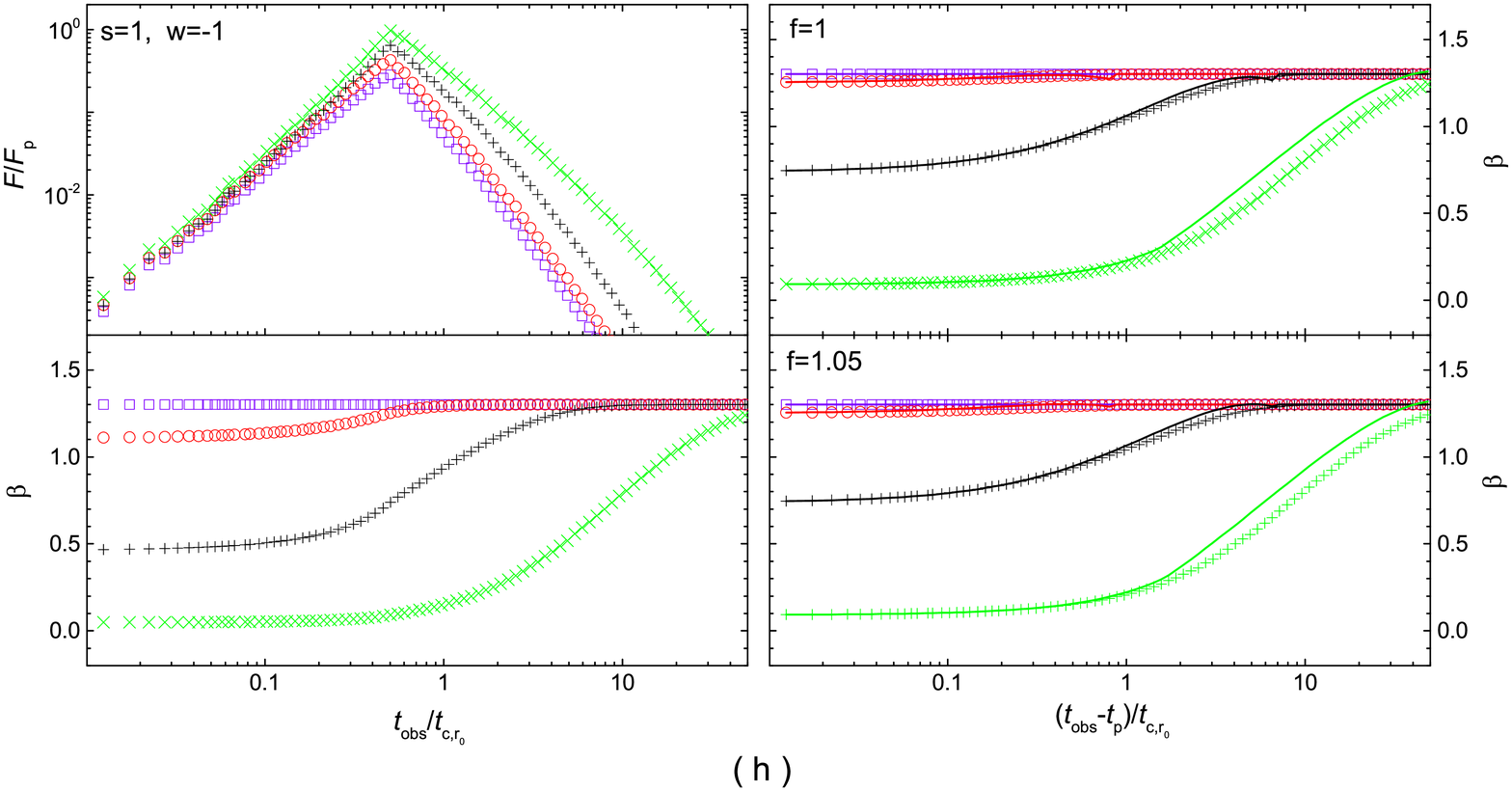}
\caption{\emph{Continued}
}
\end{figure}
 \addtocounter{figure}{-1}
\begin{figure}
\plotone{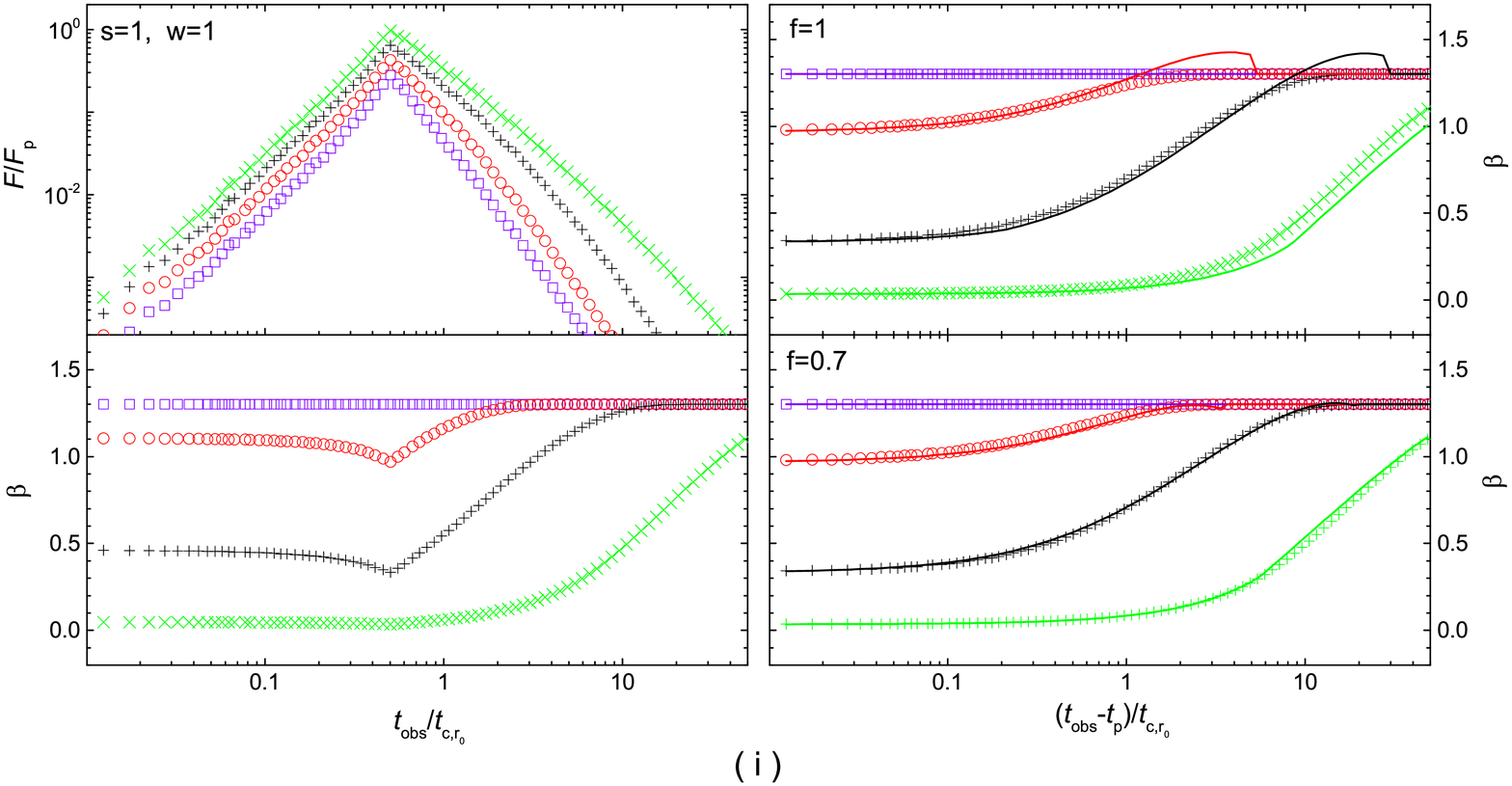}
\caption{\emph{Continued}
}
\end{figure}

\clearpage
\begin{figure}
\plotone{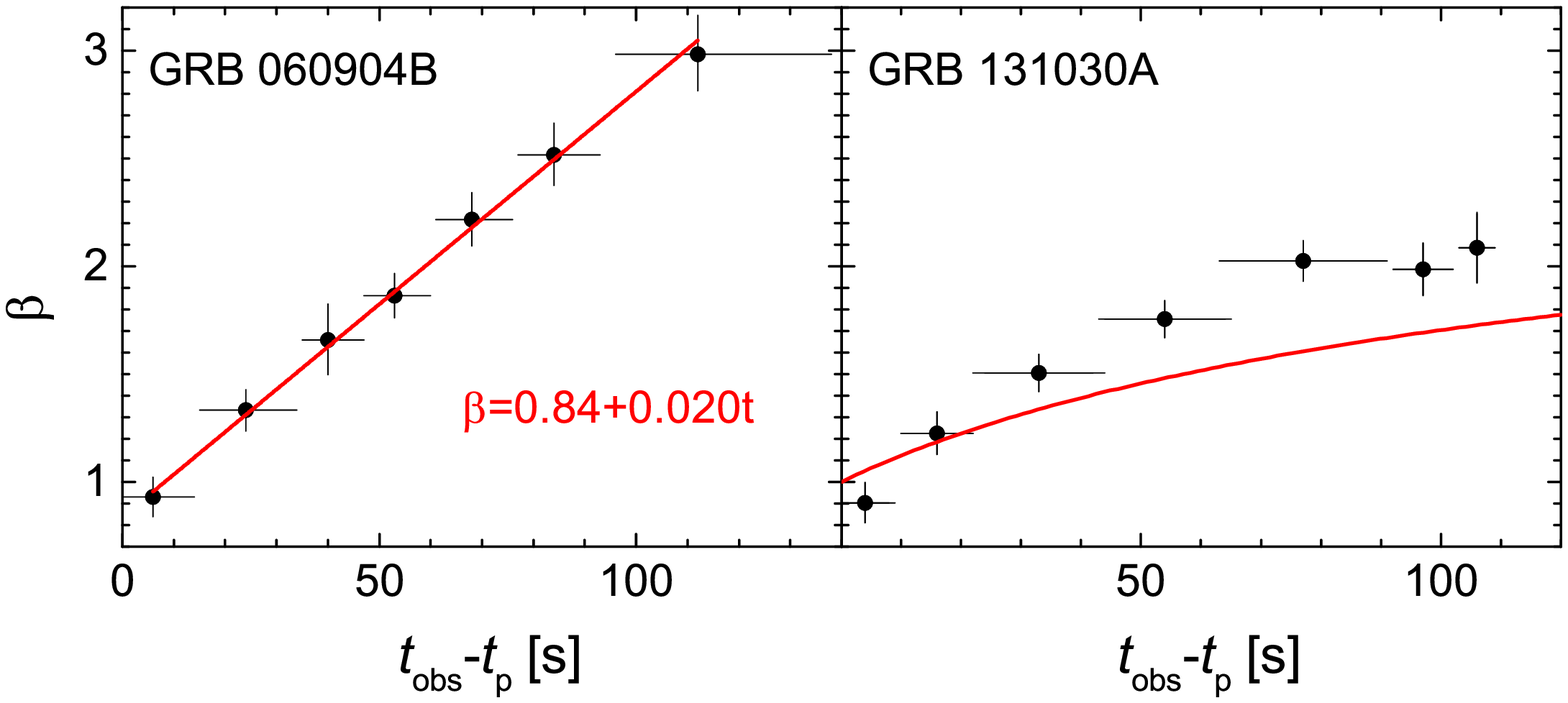}
\caption{Confront our analytical formulas of spectral evolution with observations, where $t_p$ is the peak time of our studying flares.
}
\end{figure}
\end{document}